\documentclass[twocolumn,showpacs,preprintnumbers,amsmath,amssymb,superscriptaddress,prl,longbibliography]{revtex4-2}

\usepackage{graphicx}
\usepackage{dcolumn}
\usepackage{bm}
\usepackage{graphics}
\usepackage{subfigure}
\usepackage{graphicx}
\usepackage{epsfig}
\usepackage{epstopdf}
\usepackage{xcolor}
\usepackage{multirow}
\usepackage{mathtools}

\allowdisplaybreaks

\begin{document}

\title{Nonrelativistic Conformal Invariance in Mesoscopic Two-Dimensional Fermi Gases}

\author{Viktor Bekassy}
\email{bekassy@student.chalmers.se}
\affiliation{Department of Physics, Chalmers University of Technology, 41296 Gothenburg, Sweden}

\author{Johannes Hofmann}
\email{johannes.hofmann@physics.gu.se}
\affiliation{Department of Physics, Gothenburg University, 41296 Gothenburg, Sweden}

\date{\today}

\begin{abstract}
Two-dimensional Fermi gases with universal short-range interactions are known to exhibit a quantum anomaly, where a classical scale and conformal invariance is broken by quantum effects at strong coupling. We argue that in a quasi two-dimensional geometry, a conformal window remains at weak interactions. Using degenerate perturbation theory, we verify the conformal symmetry by computing the energy spectrum of mesoscopic particle ensembles in a harmonic trap, which separates into conformal towers formed by so-called primary states and their center-of-mass and breathing-mode excitations, the latter having excitation energies at precisely twice the harmonic oscillator energy. In addition, using Metropolis importance sampling, we compute the hyperradial distribution function of the many-body wave functions, which are predicted by the conformal symmetry in closed analytical form. The weakly interacting Fermi gas constitutes a system where the nonrelativistic conformal symmetry can be revealed using elementary methods, and our results are testable in current experiments on mesoscopic Fermi gases.
\end{abstract}

\maketitle

Scale invariance arises in many areas of condensed matter physics, for example, near second order phase transitions~\cite{cardy02,amit06}. For certain interacting many-body systems, a scale symmetry may even exist generically without fine-tuning to a transition point~\cite{castin12,zwerger16,zwerger21}. The prominent example is the unitary Fermi gas in atomic physics~\cite{bloch08}, a nonrelativistic system for which a rescaling of time and position coordinates by \mbox{$(t,{\bf x}) \to (\lambda^2t, \lambda{\bf x})$} leaves the action invariant, changing the Hamiltonian by \mbox{$H \to H/\lambda^2$}; the symmetry implies, for example, a homogeneous equation of state~\cite{ku12,vanhoucke12} and a vanishing bulk viscosity~\cite{son07,schaefer12}. Even in a harmonic trap---a generic confining potential that explicitly breaks scale invariance---the properties of the gas are still constrained. This follows since scale invariance implies an additional symmetry under special conformal transformations \mbox{$(t,{\bf x}) \to (t, {\bf x})/(1+\lambda t)$}~\cite{mehen00,son06}, the generator $C$ of which takes the same form as a harmonic oscillator potential. Hence, the Hamiltonian $H_\omega$ of a trapped system is part of a nonrelativistic conformal symmetry algebra formed by $H$, $C$, and the generator of dilatations, $D$. [Formally, $D = - i X \nabla_X$ and $C = X^2/2$ with $X = ({\bf r}_{1}, {\bf r}_{2}, \ldots)$ a vector of particle coordinates, such that $H_\omega = H + m\omega^2 C$ with $m$ the particle mass and $\omega$ the trap frequency.] The conformal symmetry then implies a one-to-one correspondence between free-space eigenstates at zero energy and certain states in a harmonic trap called primary states~\cite{werner06,nishida07}. This mapping is now applied beyond cold atom physics to describe nuclear reactions in a conformal window~\cite{hammer21,schaefer21}. The symmetry also generates the spectrum of remaining nonprimary states through the ladder operator~\cite{pitaevskii97,castin04,werner06,castin12}
\begin{equation} \label{eq:defL}
    L^\dagger= - i D + \frac{H_\omega}{\hbar\omega} - \frac{C}{\ell_{\rm ho}^2} ,
\end{equation}
where \mbox{$\ell_{\rm ho} = \sqrt{\hbar/m\omega}$} is the oscillator length,
which acting on a primary state generates a breathing mode  with excitation energy $2 \hbar \omega$. The microscopic origin of this precise level spacing is linked to a separability of the many-body wave function into a hyperradial part that depends only on the modulus of the vector $X$~\cite{werner06,castin12}.

A special situation arises for atoms confined in two dimensions (2D), which are described by the Hamiltonian
\begin{equation}\label{eq:hamilton}
H_\omega = \sum_{i\sigma} \biggl( - \frac{1}{2} \nabla^2_{i\sigma} + \frac{r_{i\sigma}^2}{2} \biggr) + g \sum_{ij} \delta^{(2)}({\bf r}_{i\uparrow} - {\bf r}_{j\downarrow}) ,
\end{equation}
here written for two-component fermions with spin projection \mbox{$\sigma=\uparrow,\downarrow$}. Throughout the Letter, we use dimensionless units with \mbox{$\hbar \omega = 1$} and \mbox{$\ell_{\rm ho} = 1$}. The last term describes a universal short-range interaction with dimensionless coupling strength~$g$. Because of the homogeneity of the delta function, \mbox{$\delta^{(2)}(\lambda {\bf r}) = \lambda^{-2} \delta^{(2)}({\bf r})$}, at first sight this Hamiltonian is scale invariant. However, a delta-function interaction in two dimensions is not well defined and requires renormalization, such that $g$ is replaced by a scale-dependent ``running'' coupling \mbox{$g(\kappa) = 2\pi/\ln(1/\kappa a_2)$}~\cite{zwerger21} that depends on a 2D scattering length $a_2$ and a characteristic wave number $\kappa$ (for example, the Fermi momentum or the inverse thermal wavelength), which breaks scale invariance. This breaking of a classical symmetry by quantum fluctuations is known as a quantum anomaly~\cite{hofmann12,olhshanii10}. However, in a quasi-2D geometry with particles in the lowest state of a transverse harmonic potential with oscillator length $l_z$, the scattering length $a_2 \sim l_z \exp[-\sqrt{\pi/2} (l_z/a)]$ is an exponentially small function of the 3D scattering length $a$ in the generic situation where \mbox{$0 < a \ll l_z$}~\cite{petrov01,bloch08,zwerger16,zwerger21}. Scaling violations are then negligible, and the gas is described  by a constant (scale-invariant) interaction with strength \mbox{$g = \sqrt{8\pi} a/l_z$}~\cite{bloch08,zwerger16,hofmann21}. This is the generic situation in 2D Bose gases~\cite{chevy02,hung11,desbuquois14,saintjalm19,zou21}, and it corresponds to an easily accessed weak-interaction regime for 2D Fermi gases~\cite{froehlich11,sommer12}. Hence, while much of the discussion for Fermi gases is focussed on the quantum anomaly at stronger interactions~\cite{vogt12,gao12,chafin13,peppler18,holten18,drut18,mulkerin18,daza18,hu19,yin20}, there still exists a conformal window at small coupling.

In this Letter, we confirm and study the conformal invariance in a weakly interacting 2D Fermi gas. We focus on mesoscopic systems with a small particle number, which are in the quasi-2D regime, and describe the gas to leading linear order in the interaction strength~$g$ by means of (degenerate) perturbation theory. At this order, scale invariance is exact, with logarithmic corrections only entering at higher order: Indeed, experimental signatures of scale invariance breaking---such as a shift in the breathing mode frequency~\cite{hofmann12}, logarithmic corrections to the rf spectrum~\cite{langmack12}, or a finite bulk viscosity~\cite{hofmann20,enss19,nishida19}---only start at second order in the interaction parameter $g(\kappa)$. Moreover, on a formal level, the quantum anomaly is manifest in the commutator between $D$ and $H$, which reads $[D,H]= 2iH + i {\cal I}/2\pi$~\cite{hofmann12}. The operator ${\cal I}$ violating scale invariance is the Tan contact that parametrizes universal short-range correlations~\cite{tan08a,tan08b,tan08c,braaten08}, and its expectation value, too, starts at second order~\cite{bertaina13,anderson15,rammelmueller16,hofmann21}. In addition, although corrections to scale invariance at higher orders are expected in principle, they can be quite small~\cite{taylor12}, and we expect the conformal window to extend beyond the range of validity of first order perturbation theory. To the best of our knowledge, this provides the only setup where the nonrelativistic conformal symmetry can be verified exactly by elementary means in an interacting quantum system. Moreover, the results presented here should be observable in current experiments on interacting few-body 2D Fermi systems~\cite{serwane11,zuern13,bayha20,holten21a,holten21b}.

We begin with the Hamiltonian~\eqref{eq:hamilton} in occupation-number representation
\begin{equation} \label{eq:Ham}
H_\omega = \sum_{j,\sigma}\epsilon_{j}c^{\dagger}_{j\sigma}c_{j\sigma}^{} + g \sum_{ijkl}w_{ijkl}c^{\dagger}_{i\uparrow}c^{\dagger}_{j\downarrow}c_{k\downarrow}^{}c_{l\uparrow}^{}.
\end{equation}
Here, $c_{j\sigma}^{\dagger}$ creates a fermion with spin projection \mbox{$\sigma=\uparrow,\downarrow$} in a single-particle state \mbox{$j=\{n_j,m_j\}$} with energy \mbox{$\epsilon_j=2n_j+|m_j|+1$}, where $n_j$ is the radial quantum number and $m_j$ the angular momentum projection. Moreover, $w_{ijkl}=\int d^2r \phi_i^{*}\phi_j^{*}\phi_k\phi_l $ is the overlap integral of harmonic oscillator wave functions, $\phi_j(z,\bar{z}) = \sqrt{n_j!/\pi (n_j+|m_j|)!} z^{m_j} e^{-\bar{z}z/2} L_{n_j}^{|m_j|}(\bar{z}z)$, where $L_{n}^{|m|}$ is the associated Laguerre polynomial and we use complex particle positions \mbox{$z = x+iy$}. Single-particle states with energy $\ell+1$ are $\ell+1$-fold degenerate with angular momentum $m=-\ell,-\ell+2,\ldots,\ell$. Throughout the Letter, we consider $N$-particle configurations with an equal number of both spin types for even~$N$, and one excess spin for odd $N$. Without interactions, the ground state is obtained by successively populating the lowest single-particle levels with particles of both spins. Unless there is a ``magic'' number of particles, for which all states at a given energy are all either fully occupied or empty (this is the case for $N=2,6,12,20,30,42, \ldots$), the ground state is degenerate. So are all excited-state configurations, which have integer excitation energies and are obtained by populating higher single-particle levels. This degeneracy is lifted when interactions are taken into account. To leading order in degenerate perturbation theory, we collect all states $\{|\Psi_m\rangle\}$ with equal noninteracting energy, and diagonalize the Hamiltonian matrix~\cite{sakurai94,gottfried03},
\begin{equation}\label{eq:matrixelement}
H_{mn} = \langle \Psi_m | H_\omega | \Psi_n \rangle ,
\end{equation}
which gives the energy eigenvalues
\begin{equation}
E_N=E_N^{(0)}+E_N^{(1)} \label{eq:perturbation}
\end{equation}
with $E_N^{(0)}$ the noninteracting energy and \mbox{$E_N^{(1)} \sim {\cal O}(g)$} the interaction energy. We also determine the angular momentum projection $M$ and the total spin eigenvalue $S(S+1)$ for each state. Note that scale invariance at leading order in perturbation theory follows directly from the homogeneity of the delta potential in the matrix element~\eqref{eq:matrixelement}. The second-order correction to the energy, by contrast, includes a divergent summation over internal states~\cite{sakurai94,gottfried03}, leading to a cutoff dependence that violates scale symmetry.

%++++++++++++++++++++++++++++++++++++++++
\begin{figure}[t!]
\subfigure{\includegraphics[scale=0.89]{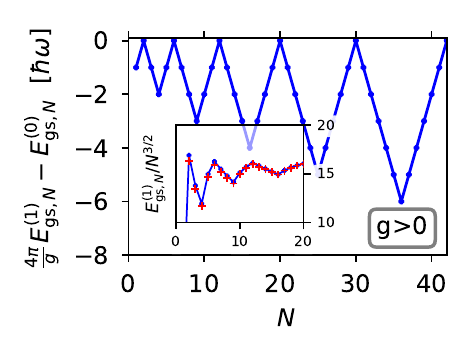}\label{fig:1a}}
\subfigure{\includegraphics[scale=0.89]{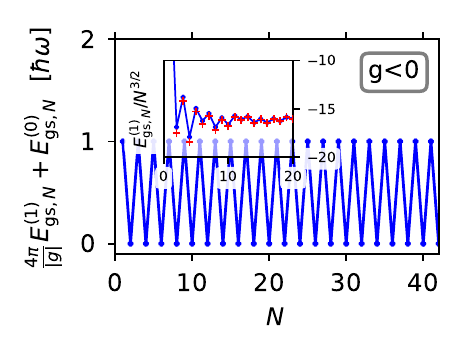}\label{fig:1b}}
\caption{
Beyond-mean-field contribution to the ground state energy in a harmonic trap as a function of particle number for (a) repulsive and (b) attractive interactions. Inset: comparison with exact diagonalization results for $|g|=0.3$ (red crosses, data from Ref.~\cite{rontani09}).
}
\label{fig:1}
\end{figure}
%++++++++++++++++++++++++++++++++++++++++

We obtain a simple analytical result for the ground state energy,
\begin{equation} \label{eq:ensp}
E_{{\rm gs},N}^{(1)} = \frac{g}{4\pi} \, E_{{\rm gs},N}^{(0)} - \frac{g}{2\pi} \, S_{{\rm gs},N} ,
\end{equation}
where $S_{{\rm gs},N}$ is the total spin of the ground state, which is determined by the particles in the valence shell. The first term describes a mean-field shift, where the energy of occupied levels is changed by a factor \mbox{$(1+g/4\pi)$}. The beyond-mean-field contribution in the second term is shown in Fig.~\ref{fig:1}. For repulsive interactions [Fig.~\ref{fig:1a}], the ground state has maximal total spin, corresponding to the largest possible number of unpaired fermions in the valence shell (the total spin is extremal for half-filled shells, $N=4,9,16,25,36,49,\ldots$). This is an example of Hund's rule~\cite{landau91}. By contrast, for attractive interactions [Fig.~\ref{fig:1b}], the total spin of an even-parity configuration is zero, \mbox{$S=0$}, while for an odd-parity configuration, it is \mbox{$S=1/2$}. This is known as the parity effect~\cite{hofmann16,hofmann17}, where odd-parity states have an excess energy compared to their even-parity neighbors. The inset of Fig.~\ref{fig:1} compares the result~\eqref{eq:ensp} (blue lines) with numerical exact-diagonalization calculations (red crosses) for $|g| = 0.3$~\cite{rontani09}, which are in good agreement, indicating that perturbation theory is valid at these interactions.

\begin{figure}[t!]
\includegraphics[scale=0.33]{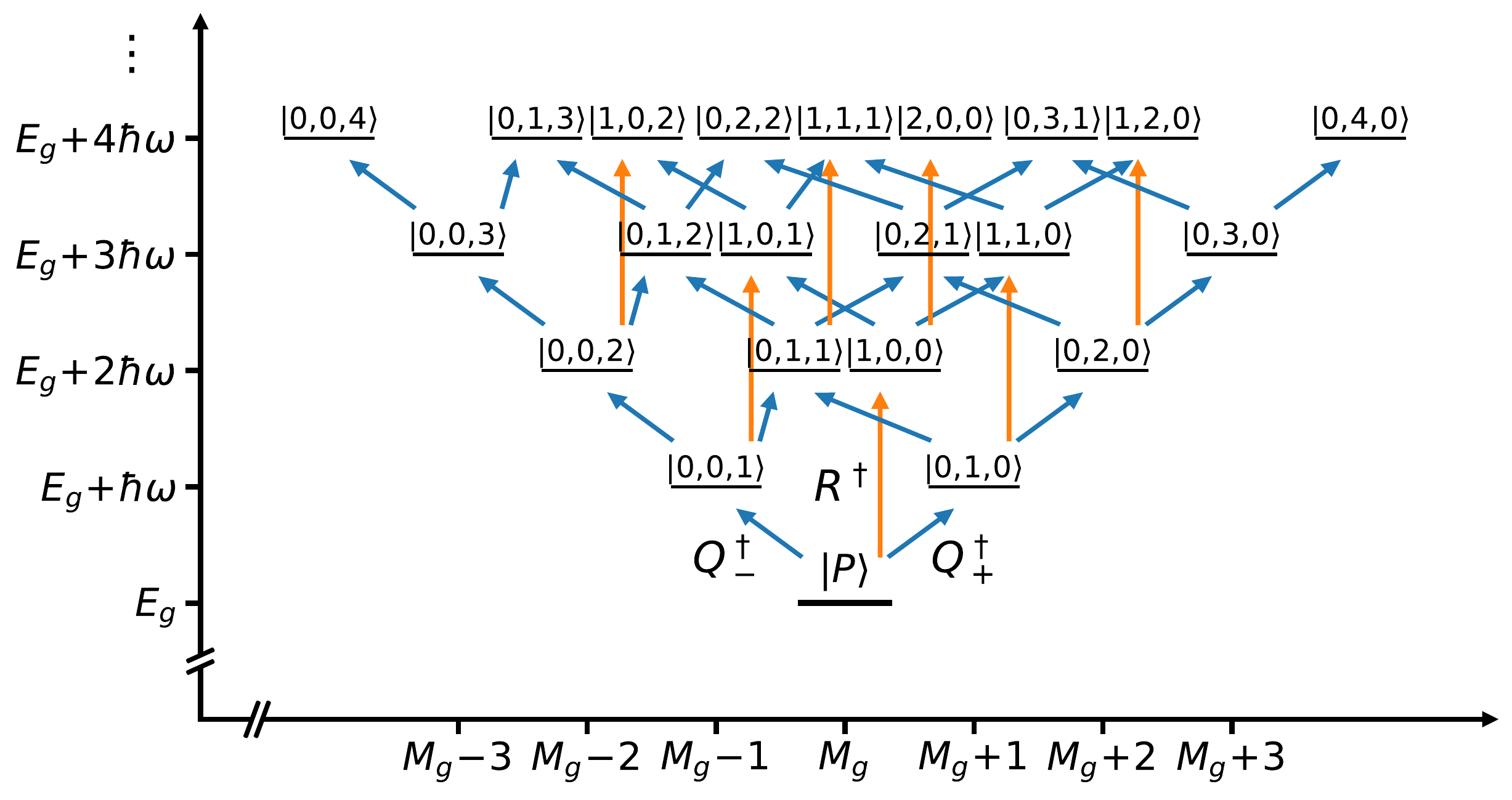}
\caption{
Conformal tower of states created from a primary state $|P\rangle$, ordered by energy and angular momentum. Nonprimary states are center-of-mass excitations, which are created by the operators $Q_\pm^\dagger$ (blue arrows) that increase the energy by $\hbar \omega$ and the angular momentum by $\pm 1$, and internal breathing mode excitations, which are created by $R^\dagger$ (orange arrows) which increases the internal energy by $2\hbar \omega$ without changing the angular momentum.
}
\label{fig:2}
\end{figure}

To discuss the excited state spectrum in a 2D harmonic trap, we introduce two spectrum-generating operators $Q_\pm^\dagger$ in addition to the operator $L^\dagger$ that generates the breathing mode. They create center-of-mass (c.m.) excitations and are defined as
\begin{eqnarray} \label{eq:defQ}
 Q_+^\dagger &= \sum_{i\sigma}  \bigl( - i \sqrt{2} \frac{\partial}{\partial \bar{z}_{i\sigma}} + i \frac{1}{\sqrt{2}} z_{i\sigma} \bigr) , \\
 Q_-^\dagger &= \sum_{i\sigma}  \bigl( - i \sqrt{2} \frac{\partial}{\partial z_{i\sigma}} + i \frac{1}{\sqrt{2}} \bar{z}_{i\sigma} \bigr) .
\end{eqnarray}
They obey the nonzero commutation relations \mbox{$[Q_\pm, Q_\pm^\dagger] = 2N$} and \mbox{$[H,Q_\pm^\dagger] = Q_\pm^\dagger$}, which are independent of the interaction potential and thus hold irrespective of scale invariance. Acting with $Q_\pm^\dagger$ on an eigenstate with energy $E$ and angular momentum $M$ creates a state with $E+1$ and $M\pm 1$. In general, however, breathing mode and c.m. excitations are not independent, which follows from the nonzero commutators $[L^\dagger, Q_\pm] = - 2 Q_\mp^\dagger$ and $[L, Q_\pm^\dagger] = 2 Q_\mp$. This is also apparent in an occupation-number representation, where $Q_\pm^\dagger$ are single-particle operators that transfer occupied states with energy $\ell$ and angular momentum $m_\ell$ to empty levels with $\ell+1$ and $m_\ell \pm 1$. To leading order in perturbation theory, $L^\dagger$ is also a single-particle operator that creates single-particle excitations by $2$ without a change in angular momentum. States generated by $L^\dagger$ and $Q_+^\dagger Q_-^\dagger$ thus have finite overlap. In order to disentangle breathing modes and c.m. excitations, following~\cite{werner06,castin12,moroz12} we introduce the operator
\begin{equation}
   R^\dagger 
   = L^\dagger - \frac{1}{2 N} \bigl( Q_{+}^{\dagger}Q_{-}^{\dagger} + Q_{-}^{\dagger}Q_{+}^{\dagger} \bigr) ,
\end{equation}
which commutes with $Q_\pm^\dagger$ since it only acts on an internal hyperradius \mbox{$\tilde{R} = \sqrt{\sum_{i\sigma} |{\bf r}_{i\sigma} - {\bf C}|^2}$}, with ${\bf C}$ the c.m. position. $R^\dagger$ thus generates {\it internal} breathing modes, again with excitation energy $2$ (which follows from $[H, R^\dagger] = + 2 R^\dagger$). Repeated c.m. and breathing mode excitations then give the orthogonal set of excited states
\begin{equation}
|a,b,c\rangle_P = (R^\dagger)^a (Q_+^\dagger)^b (Q_-^\dagger)^c |P\rangle ,
\end{equation}
where the so-called primary state $|P\rangle$ that forms the ground step is annihilated by $R$, $Q_+$, and $Q_-$. This is illustrated in Fig.~\ref{fig:2}. Denoting the energy and angular momentum of $|P\rangle$ by $E_g$ and $M_g$, the excited state has energy $E_{a,b,c} = E_g + (2 a + b + c)$ (with  internal energy $E_{int} = E_g + 2 a-1$ and c.m. energy $E_{cm} = b+c+1$) and angular momentum $M_{a,b,c} = M_g + (b - c)$. The total spin is conserved. Note that there is an infinite number of primary states, and primary and nonprimary states form a complete set of the Hilbert space. States within different conformal towers are disentangled by computing the expectation value of the Casimir operator
\begin{equation}\label{eq:casimir}
    T=4\left(T_3^2 - T_1^2-T_2^2\right),
\end{equation}
which is formed from the operators
\begin{equation}
\begin{split}
&T_1=\frac{1}{4}\left(R^{\dagger}+R\right),\; T_2=\frac{1}{4i}\left(R^{\dagger}-R\right),\;\\
&T_3=\frac{1}{2}H - \frac{1}{4N}\left(Q_+^{} Q_+^{\dagger}+Q_{-}^{\dagger} Q_{-}^{}\right) 
\end{split}
\end{equation}
that obey the nonrelativistic conformal SO(2,1) symmetry algebra $[T_1,T_2]=-iT_3,\;[T_2,T_3]=iT_1$, and $[T_3,T_1]=iT_2$. The Casimir commutes with all symmetry operators and is thus constant within each conformal tower. Evaluated for a primary state, we have $\langle P | T | P \rangle = (E_g^{(0)}-1) (E_g^{(0)}-3)$, where at this order in perturbation theory $E_g^{(0)}$ denotes the noninteracting contribution to the ground step energy of a conformal tower.

%++++++++++++++++++++++++++++++++++++++++
%++++++++++++++++++++++++++++++++++++++++

%++++++++++++++++++++++++++++++++++++++++
\begin{figure*}
\subfigure{\includegraphics[scale=0.9]{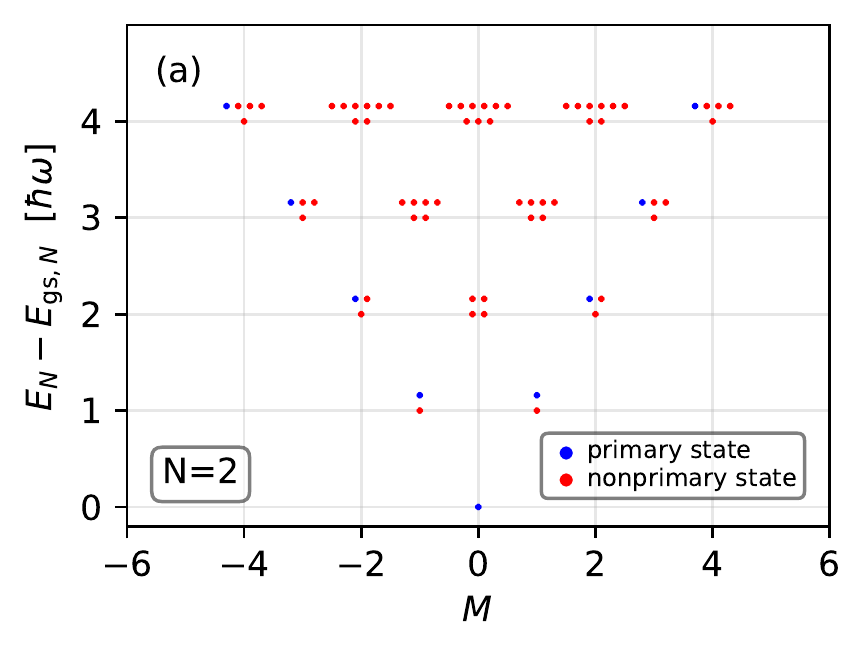}\label{fig:3a}} \qquad
\subfigure{\includegraphics[scale=0.9]{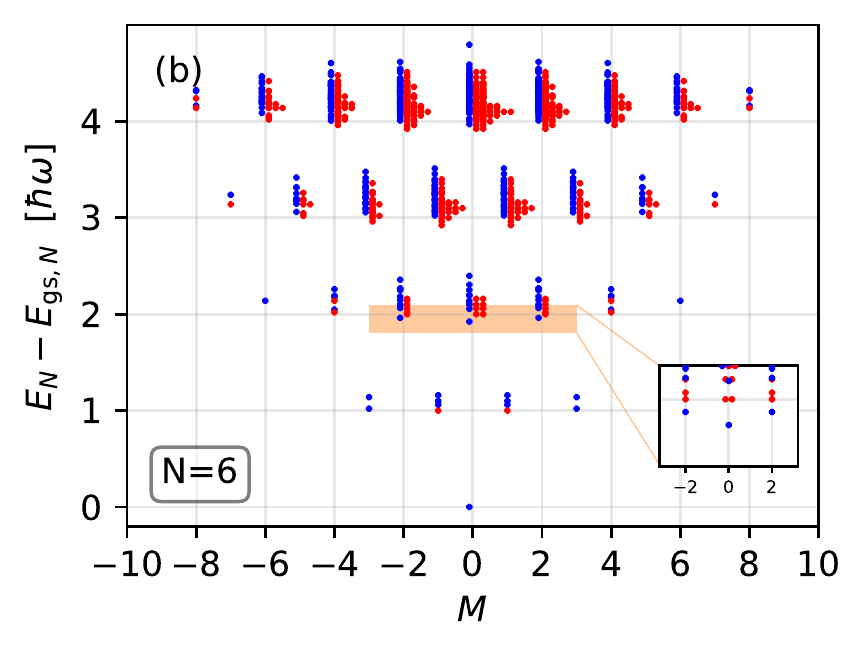}\label{fig:3b}}
\subfigure{\includegraphics[scale=0.9]{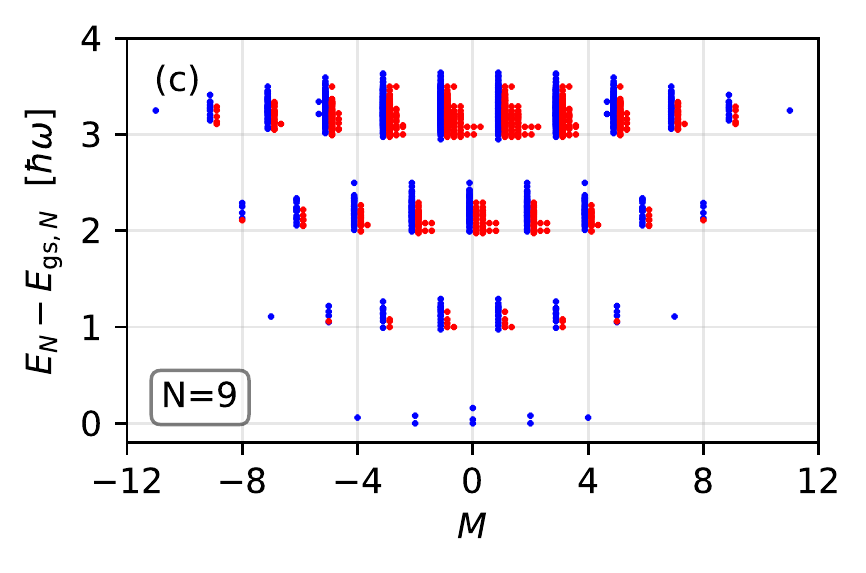}\label{fig:3c}} \qquad
\subfigure{\includegraphics[scale=0.9]{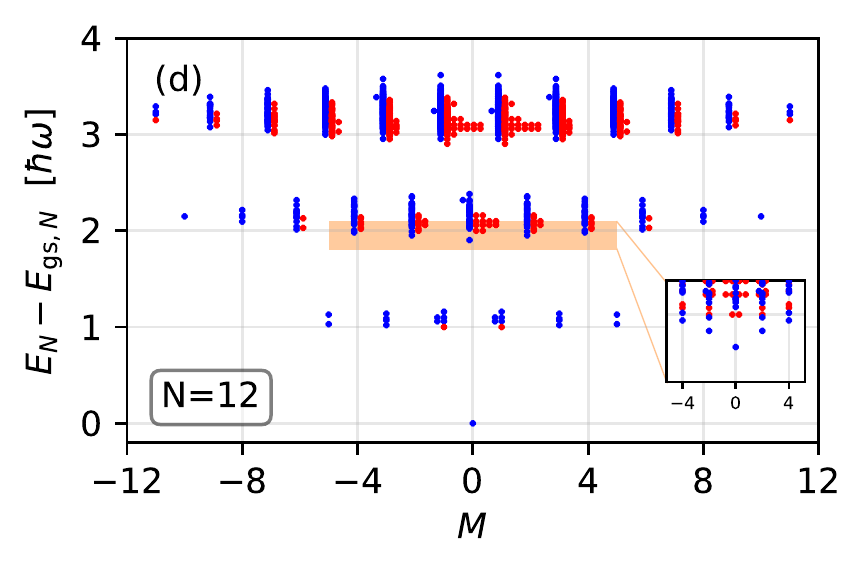}\label{fig:3d}}
\caption{
Excitations energies for $N=2,6,9$ and $12$ particles in a harmonic trap ordered by angular momentum for an attractive interaction \mbox{$g=-1$}. Blue points represent primary states and red points are nonprimary states (cf.~Fig.~\ref{fig:2}). Overlapping points are moved horizontally for clarity. Insets: Magnified spectrum near the second excitation level.
}
\label{fig:3}
\end{figure*}
%++++++++++++++++++++++++++++++++++++++++

Figure~\ref{fig:3} shows the results for the excitation energies as obtained from degenerate perturbation theory for four particle numbers $N=2,6,9$, and $12$, grouped by angular momentum. Here, $N=2,6$, and $12$ are the first three ``magic'' numbers, and \mbox{$N=9$} has a half-filled valence orbital in the ground state. We choose an attractive interaction with strength \mbox{$g=-1$}, such that the lifting of the degeneracy of noninteracting states is clearly visible, yet states remain clustered around their noninteracting excitation energies. We identify primary states and the degree of excitation of nonprimary states by computing the first integers $(a,b,c)$ for which an eigenstate is in the kernel of $R^{a+1}$, $Q_+^{b+1}$, and $Q_-^{c+1}$. In Fig.~\ref{fig:3}, we indicate primary states in blue and nonprimary states in red, where for clarity we do not differentiate different conformal towers~\cite{supplemental}. Remaining degenerate states are offset horizontally. Note that while the structure of nonprimary states is entirely dictated by the nonrelativistic conformal symmetry [cf. Fig.~\ref{fig:2}], the positions of primary states follow from our numerical calculations. In all our calculations, we verified the spectrum as predicted by the conformal symmetry. For a direct visual inspection, the spectrum is most apparent for \mbox{$N=2$}, Fig.~\ref{fig:3a}. Note that for $N=6$ and $12$, the excitation energies of several primary states at the second level is reduced compared to the noninteracting value $2$ [inset and shaded areas in Figs.~\ref{fig:3b} and ~\ref{fig:3d}]. These states contain two excitations from the ground state by one energy level, and the reduction in energy is caused by attractive interactions within the excited shell. The excitation energy of such states was studied experimentally~\cite{bayha20} and also using exact diagonalization~\cite{bjerlin16}, and our results for the lowest interaction shift \mbox{$\Delta E_6 = - 0.077 |g| = - 0.484 E_B$} and \mbox{$\Delta E_{12} = - 0.097 |g| = - 0.608 E_B$}, where $E_B = |E_{gs,N=2}^{(0)}| = |g|/2\pi$ is the two-body bound state energy [cf. Eq.~\eqref{eq:ensp}], are in agreement. For ground states with partially filled shells (such as $N=9$), a negative shift of the excitation energy exists already at the first level [cf. Fig.~\ref{fig:3}(c)].

\begin{figure}[t!]
\includegraphics[scale=0.9]{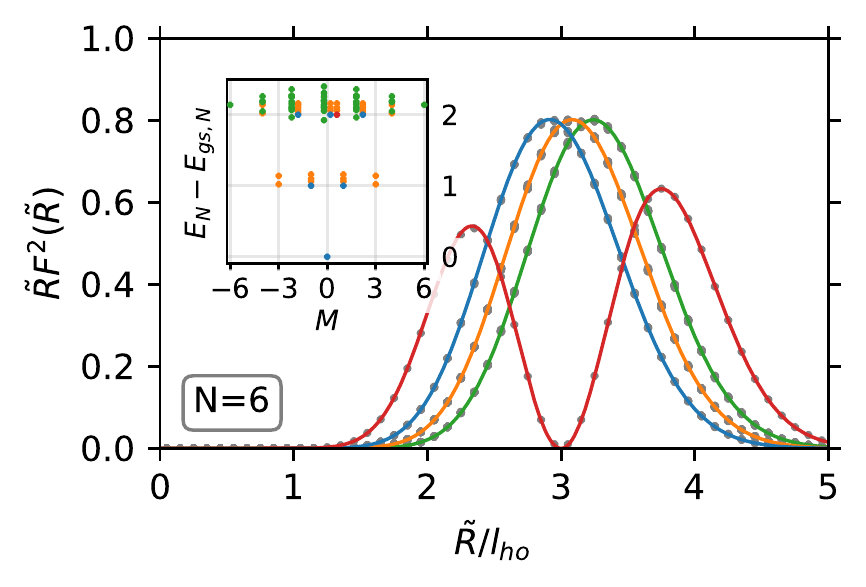}
\caption{
Distribution of the internal hyperradius $\tilde{R}$ for the $77$ lowest eigenstates of \mbox{$N=6$} particles. Gray points are results of the Monte Carlo sampling of the wave function, and continuous lines show the analytical result~\eqref{eq:hyperradius}. The inset shows the same energy spectrum as Fig.~\ref{fig:3b} with a color coding that matches the distribution.
}
\label{fig:4}
\end{figure}

As discussed in the introduction, the microscopic origin of the nonrelativistic conformal symmetry is a factorization of the many-body wave function~\cite{werner06,castin12}
\begin{equation}
\Psi({\bf r}_{1\uparrow}, \ldots, {\bf r}_{1\downarrow}, \ldots) = \Psi_{{\rm c.m.}}({\bf C}) \,\frac{F(\tilde{R})}{\tilde{R}^{N-2}} \,\phi({\bf n}) ,
\end{equation}
where $\Psi_{{\rm c.m.}}({\bf C})$ is the c.m. part (which factorizes for any Galilean-invariant interaction), $F(\tilde{R})$ the internal hyperradial part, and $\phi({\bf n})$ a  hyperangular part that depends on the remaining internal coordinates \mbox{${\bf n} = ({\bf r}_{1\uparrow} - {\bf C}, \ldots, {\bf r}_{1\downarrow} - {\bf C}, \ldots)/\tilde{R}$}. For a state $|a,b,c\rangle_P$, $F(\tilde{R})$ is determined by the identity $(R)^{a+1} |a,b,c\rangle_P = 0$:
\begin{equation}
F(\tilde{R}) = \sqrt{\frac{2 a!}{\Gamma(s+a+1)}} \tilde{R}^s e^{-\tilde{R}^2/2} L_a^s(\tilde{R}^2) , \label{eq:hyperradius}
\end{equation}
where $\Gamma$ is the Gamma function, $L_a^s$ is an associated Laguerre polynomial, and $s$ parametrizes the energy of the primary state as \mbox{$E_g^{(0)} = s+1$}~\cite{supplemental}. Note that the internal hyperradial wave function only depends on the primary state energy $E_g^{(0)}$ and the number of internal breathing mode excitations $a$ with exited states having multiple nodes. It does not depend on the angular momentum $M$ or the number of c.m. excitations $b$ and $c$, which do not affect the internal dynamics. An observable consequence of the separability is that $\tilde{R} F^2(\tilde{R})$ describes the distribution of the internal hyperradius~$\tilde{R}$~\cite{blume07,castin12}. We confirm this result using Metropolis Monte Carlo sampling of the perturbative wave function \mbox{$|\Psi_{a,b,c}({\bf r}_{1\uparrow}, \ldots, {\bf r}_{1\downarrow}, \ldots)|^2$}. Figure~\ref{fig:4} shows the hyperradial distribution for the lowest $77$ states of $N=6$ particles (corresponding to the first two excitation levels), where points are numerical results and continues lines are the analytical prediction~\eqref{eq:hyperradius}. The inset in Fig.~\ref{fig:4} shows the same spectrum as Fig.~\ref{fig:3b} with a revised color coding that matches the distribution function. As is apparent from the figure, states with an equal number of internal breathing mode excitations $a$ that are derived from primary states at the same excitation level (i.e., with equal $s$) share the same hyperradial distribution. The hyperradial distribution should be observable experimentally by sampling the many-body wave function using recently developed single-atom imaging techniques~\cite{holten21a,holten21b}, thus verifying the conformal symmetry on a microscopy level, with deviations from our predictions (for example, at stronger interactions or for deformed or rotating traps) a signature of anomalous or explicit symmetry breaking. More broadly, the mesoscopic 2D Fermi gas constitutes an experimentally relevant toy model in which the conformal symmetry can be studied exactly using elementary techniques. In particular, this provides a new way to study conformal nonequilibrium dynamics~\cite{bamler15,maki18,maki19,saintjalm19,lv20,shi20,olshanii21}.

%++++++++++++++++++++++++++++++++++++++++
%++++++++++++++++++++++++++++++++++++++++

\begin{acknowledgments}
We thank Stellan \"Ostlund and Wilhelm Zwerger for discussions and comments. We also thank Massimo Rontani for sharing exact diagonalization data from Ref.~\cite{rontani09}. This work is supported by Vetenskapsr\aa det (Grant No. 2020-04239).
\end{acknowledgments}

\bibliography{bib_perturbation}

\end{document}

\title{Supplemental Material for ``Nonrelativistic Conformal Invariance in Mesoscopic Two-Dimensional Fermi Gases"}

\author{Viktor Bekassy}
\email{bekassy@student.chalmers.se}
\affiliation{Department of Physics, Chalmers University of Technology, 41296 Gothenburg, Sweden}

\author{Johannes Hofmann}
\email{johannes.hofmann@physics.gu.se}
\affiliation{Department of Physics, Gothenburg University, 41296 Gothenburg, Sweden}

\date{\today}
\maketitle

\section{Center-of-mass distribution}

In the main text, we discuss the factorization of the many-body wave function in a center-of-mass (c.m.) part $\Psi_{\rm c.m.}({\bf C})$ and a hyperradial part $F(\tilde{R})$; cf. Eq.~(13). The latter is determined by the nonrelativistic conformal symmetry in closed analytical form, cf. Eq.~(14), which in turns provides an analytical result for the distribution of the hyperradius $\tilde{R}$ that we confirm using Monte Carlo simulations as shown in Fig.~4. \\

Here, we present similar analytical results for the center-of-mass wavefunction $\Psi_{\rm c.m.}$. For a state $|a,b,c\rangle_P$ (defined in Eq.~(10) of the main text), $\Psi_{\rm c.m.}$ is determined by the identities $(Q_+)^{b+1} |a,b,c\rangle_P = 0$ and $(Q_-)^{c+1} |a,b,c\rangle_P = 0$, which gives for $b \geq c$
\begin{align}
\Psi_{\rm c.m.}(C) &= \sqrt{ \frac{2 N^{1 + b-c} c!}{b!}} C^{b-c} e^{-N |C|^2/2} L_{c}^{b-c}(N |C|^2) , \label{eq:comanalytic}
\end{align}
with $C$ and $\bar{C}$ as well as $b$ and $c$ exchanged for $c \geq b$. Note that, different from the internal hyperradial wavefunction $F(\tilde{R})$, this result does not depend on the internal energy of the primary state or the internal breathing mode excitation and only depends on the center-of-mass excitations described by the quantum numbers $b$ and $c$. States derived from different primary states that have an equal number of center-of-mass excitations and the same total angular momentum imparted on them will thus have the same center-of-mass distribution. \\

\begin{figure*}
\includegraphics[scale=1.]{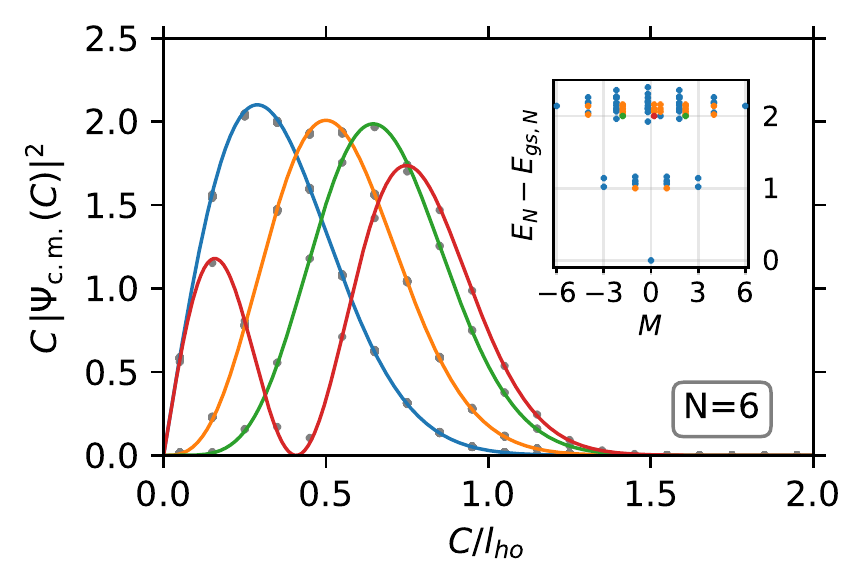}\label{fig:3a}
\caption{
Distribution of the center-of-mass positions $C$ for the $77$ lowest eigenstates of \mbox{$N=6$} particles. Gray points are results of the Monte Carlo sampling of the wave function, and continuous lines show the analytical result~\eqref{eq:comanalytic}. The inset shows the same energy spectrum as in Fig.~3 of the main paper with a color coding that matches the distribution.
}
\label{fig:S1}
\end{figure*}

As for the hyperradial distribution, these analytical predictions can be verified using Monte Carlo simulations. Figure~\ref{fig:S1} shows the distribution of the center-of-mass coordinate $C$ sampled for the same $77$ lowest-lying states of the $N=6$ particle system as in Fig.~4 of the main text, where the inset shows the energy spectrum for these state with a color coding that matches the distribution. It is important to note that (unlike for the hyperradial distribution) the data collapse onto four different scaling curves is not related to the conformal symmetry at this order in perturbation theory, but follows generally  for the decoupled center-of-mass dynamics in a Galilean-invariant system. 

%++++++++++++++++++++++++++++++++++++++++
%++++++++++++++++++++++++++++++++++++++++

\newpage

\section{Extended data}

In this section, we provide extended data and plots for the lowest-lying states of $N=2,6,9$, and $12$ particles. \\

Figures show the energy spectrum in the same arrangement as in Fig.~3 of the main text. Panels have a different color coding to indicate (a) primary and nonprimary states, (b) the value of the Casmir, (c) the total spin, (d) the number of internal breathing modes $a$ and center-of-mass excitations (e) $b$ and (f) $c$, respectively. The first panel (a) indicating primary and nonprimary states is the same as in the main text. \\

Tables list the primary states up to the first two excitation levels along with their energy, total spin quantum number, angular momentum, and Casimir. \\

In addition, an accompanying supplemental file in csv format, which is linked on the journal website, contains exact diagonalization data for a total of 23054 states corresponding to the first few excitations levels of $N=2-20$ particles. Each row contains a state and columns are ordered according to: particle number $N$; excitation level; angular momentum $M$; total spin quantum number $s$; leading-order ground state energy $E_N^{(0)}$; first-order correction to the ground state energy $E_N^{(1)}$; Casimir $T$; a flag indicating if the state is primary or not (1/0); number of internal breathing mode excitations $a$ generated by $R^\dagger$; number of center-of-mass excitations $c$ generated by $Q_-^\dagger$; and the number of center-of-mass excitations $b$ generated by $Q_+^\dagger$. Units are the same as in the main text.

%++++++++++++++++++++++++++++++++++++++++
\begin{figure*}[h!]
\subfigure[primary/nonprimary states]
	{\raisebox{-0.2cm}{\includegraphics[scale=0.8]{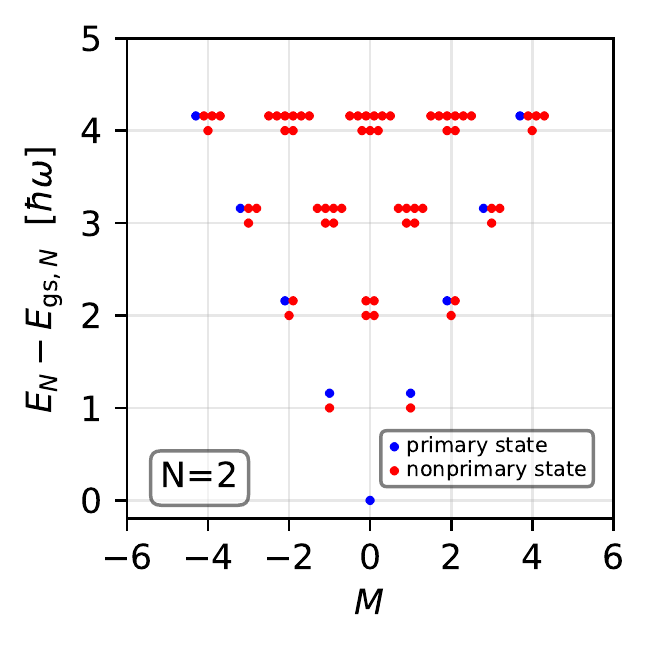}} } 
\subfigure[Casimir $\langle T \rangle$]
	{\raisebox{-0.2cm}{\includegraphics[scale=0.8]{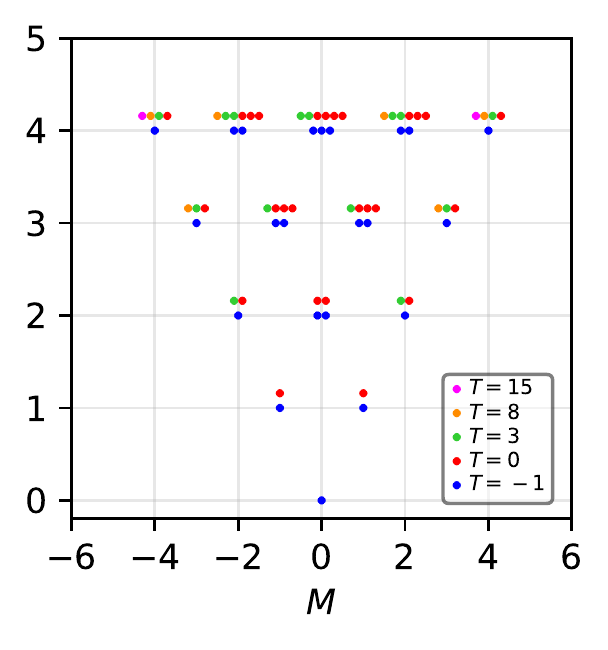}} }
\subfigure[total spin $\langle S^2 \rangle = s (s+1)$]
	{\raisebox{-0.2cm}{\includegraphics[scale=0.8]{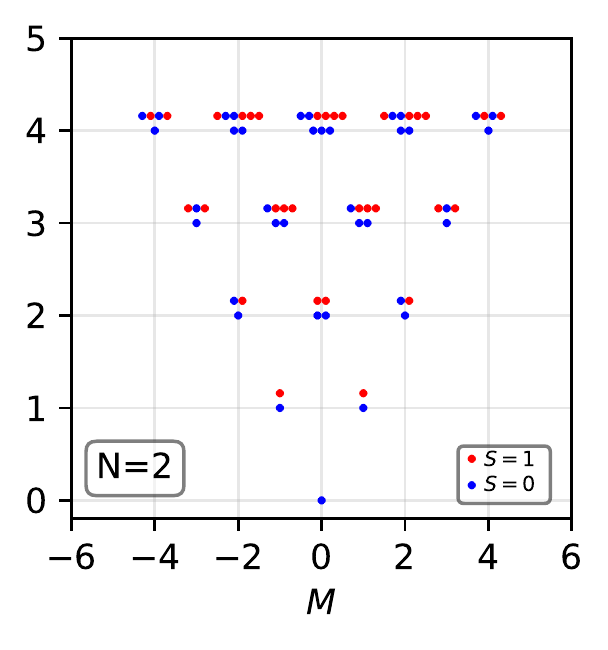}} } \\[-2ex]
\subfigure[Internal breathing modes $(R^\dagger)^a$]
	{\raisebox{-0.2cm}{\includegraphics[scale=0.8]{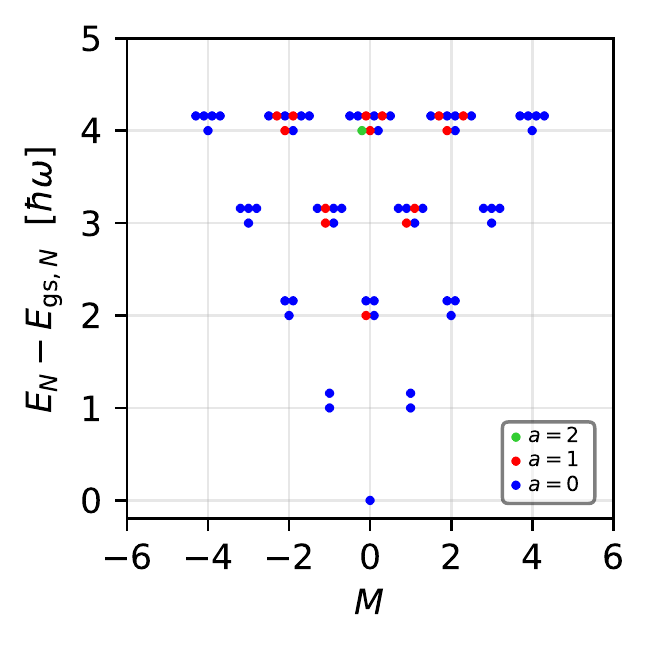}} }
\subfigure[Center-of-mass excitations $(Q_+^\dagger)^b$]
	{\raisebox{-0.2cm}{\includegraphics[scale=0.8]{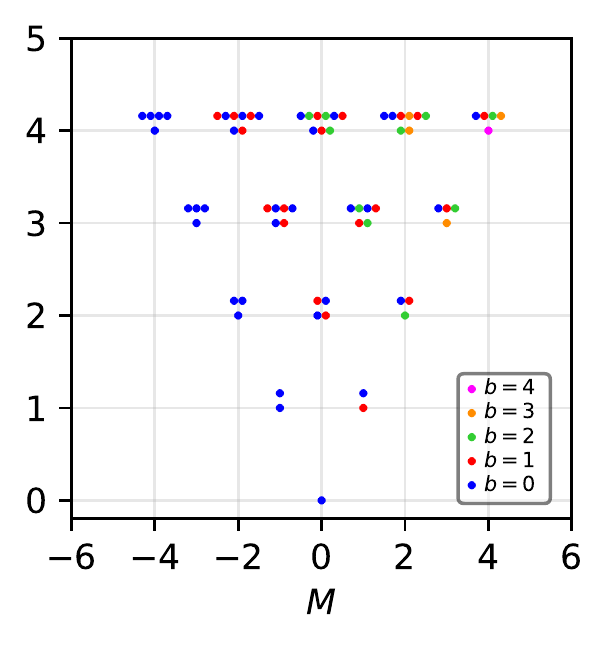}} }
\subfigure[Center-of-mass excitations $(Q_-^\dagger)^c$]
	{\raisebox{-0.2cm}{\includegraphics[scale=0.8]{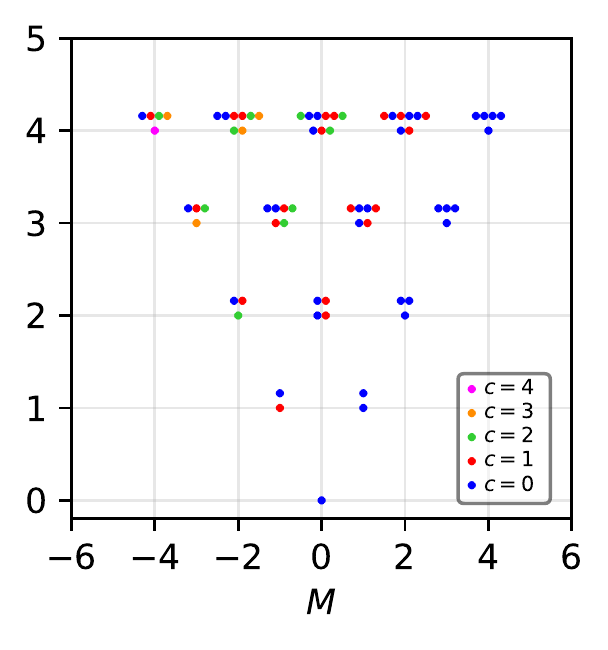}} }
\caption{
Excitations energies for $N=2$ particles in a harmonic trap ordered by angular momentum for an attractive interaction \mbox{$g=-1$}. Data points are the same for all panels, but the color coding differs: States are grouped according to (a) primary and nonprimary states, (b) Casimir operator $\langle T \rangle$, (c) total spin quantum number $s$, (d) Number of internal breathing mode excitations $a$, and (e) and (f) number of center-of-mass excitations $b$ and $c$.
}
\label{fig:S2}
\end{figure*}
%++++++++++++++++++++++++++++++++++++++++

%++++++++++++++++++++++++++++++++++++++++
\begin{figure*}[h!]
\subfigure[primary/nonprimary states]
	{\raisebox{-0.2cm}{\includegraphics[scale=0.8]{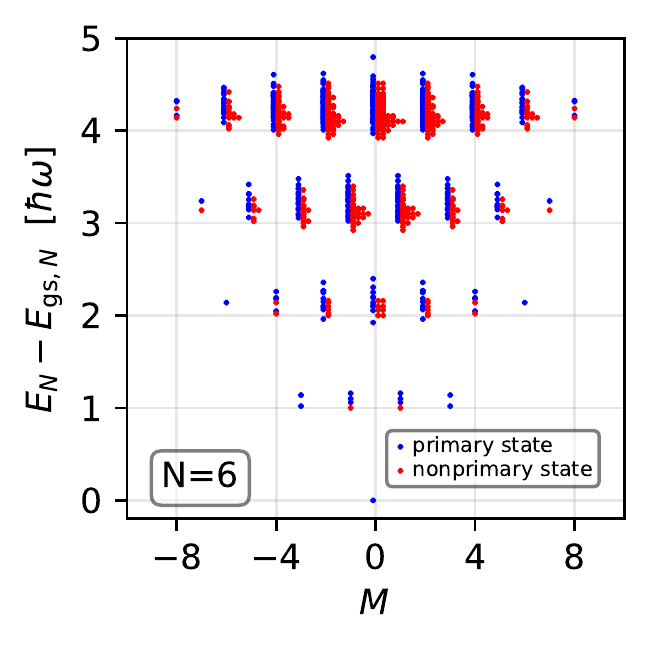}} } 
\subfigure[Casimir $\langle T \rangle$]
	{\raisebox{-0.2cm}{\includegraphics[scale=0.8]{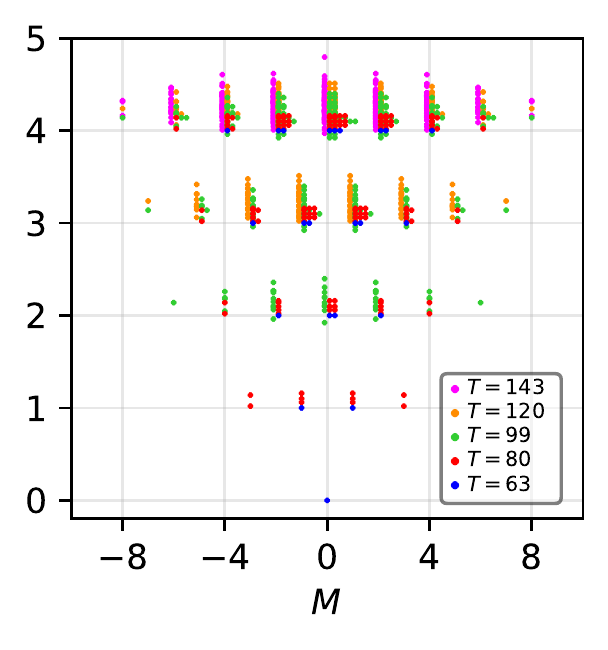}} }
\subfigure[total spin $\langle S^2 \rangle = s (s+1)$]
	{\raisebox{-0.2cm}{\includegraphics[scale=0.8]{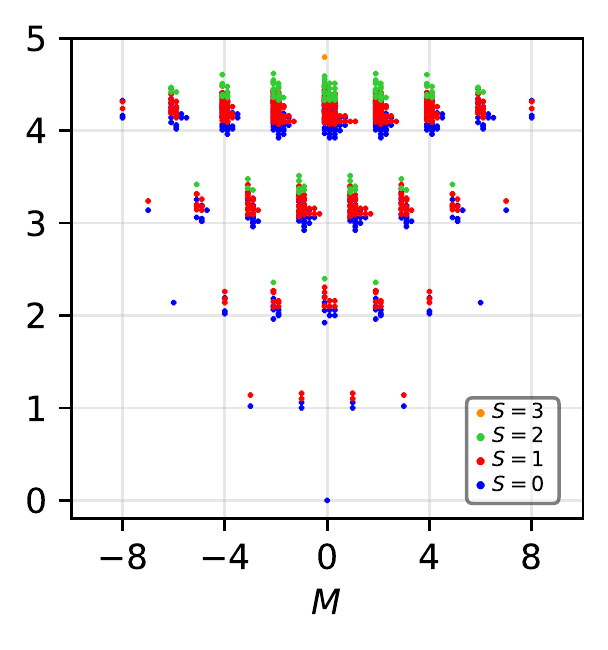}} } \\[-2ex]
\subfigure[Internal breathing modes $(R^\dagger)^a$]
	{\raisebox{-0.2cm}{\includegraphics[scale=0.8]{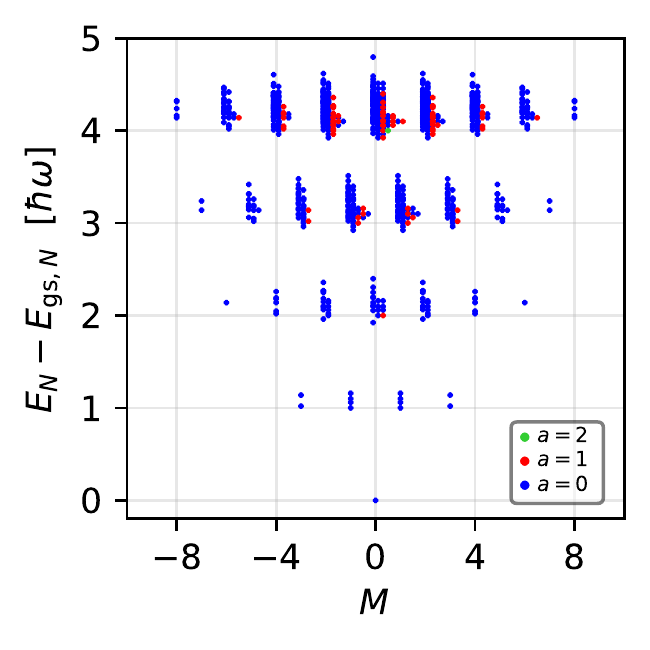}} }
\subfigure[Center-of-mass excitations $(Q_+^\dagger)^b$]
	{\raisebox{-0.2cm}{\includegraphics[scale=0.8]{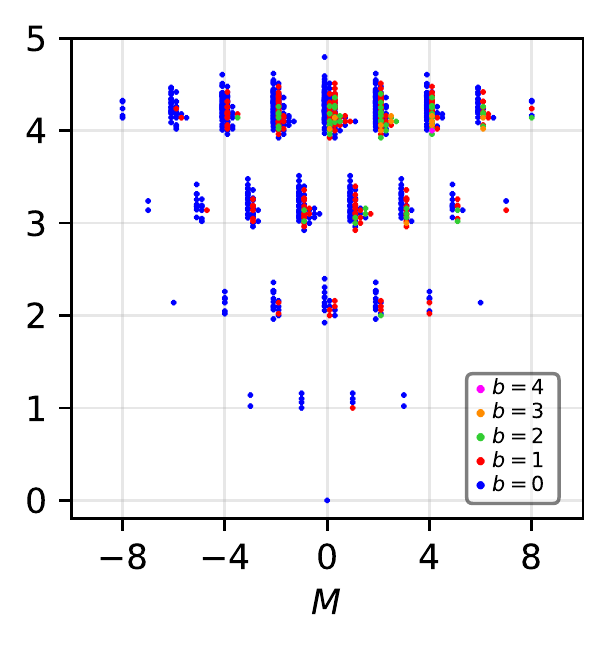}} }
\subfigure[Center-of-mass excitations $(Q_-^\dagger)^c$]
	{\raisebox{-0.2cm}{\includegraphics[scale=0.8]{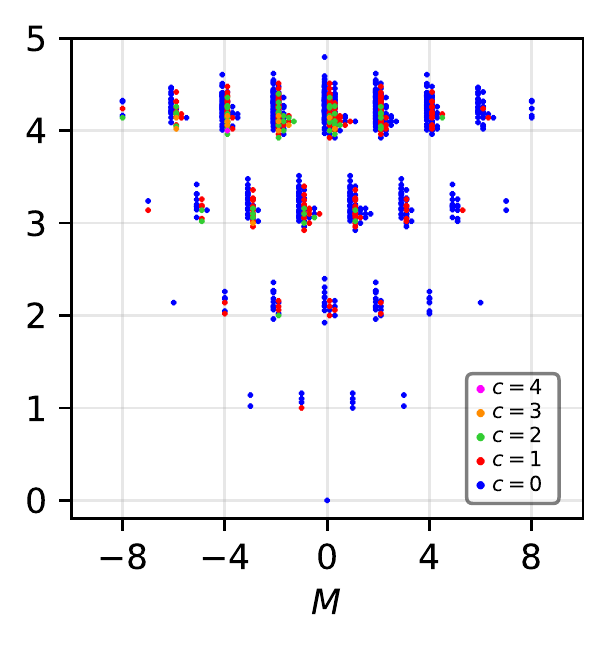}} }
\caption{
Excitations energies for $N=6$ particles in a harmonic trap ordered by angular momentum for an attractive interaction \mbox{$g=-1$}. Data points are the same for all panels, but the color coding differs: States are grouped according to (a) primary and nonprimary states, (b) Casimir operator $\langle T \rangle$, (c) total spin quantum number $s$, (d) Number of internal breathing mode excitations $a$, and (e) and (f) number of center-of-mass excitations $b$ and $c$.
}
\label{fig:S3}
\end{figure*}
%++++++++++++++++++++++++++++++++++++++++

%++++++++++++++++++++++++++++++++++++++++
\begin{figure*}[h!]
\subfigure[primary/nonprimary states]
	{\raisebox{-0.2cm}{\includegraphics[scale=0.8]{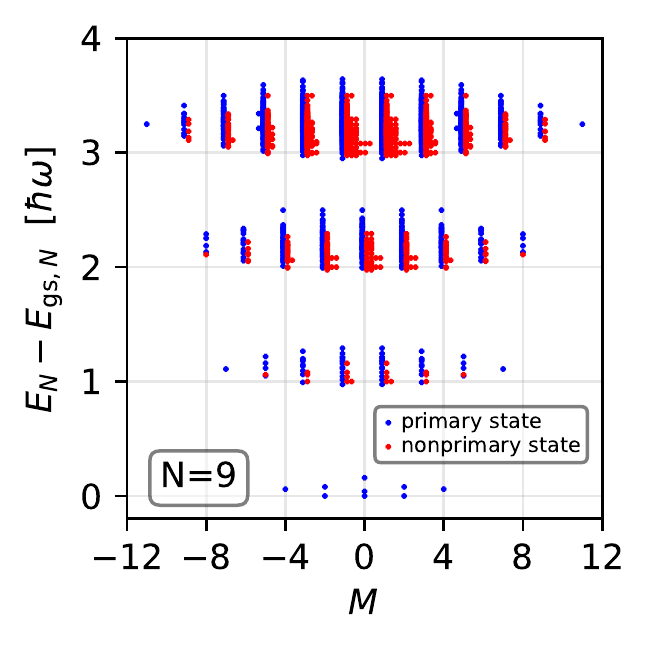}} } 
\subfigure[Casimir $\langle T \rangle$]
	{\raisebox{-0.2cm}{\includegraphics[scale=0.8]{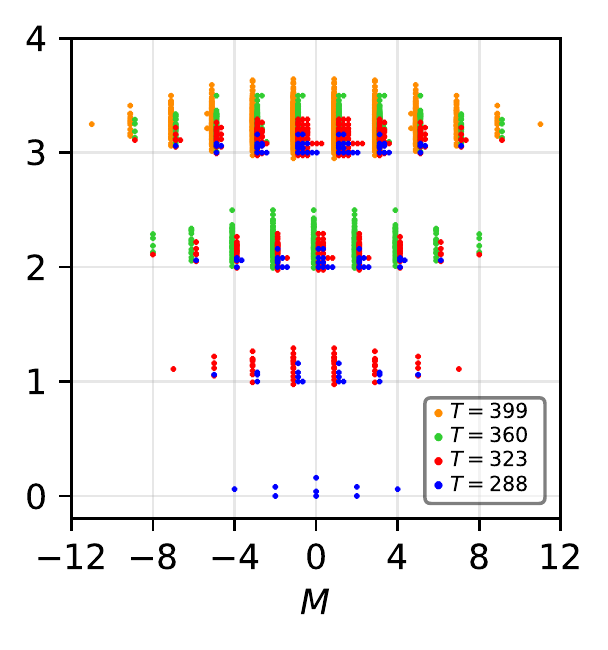}} }
\subfigure[total spin $\langle S^2 \rangle = s (s+1)$]
	{\raisebox{-0.2cm}{\includegraphics[scale=0.8]{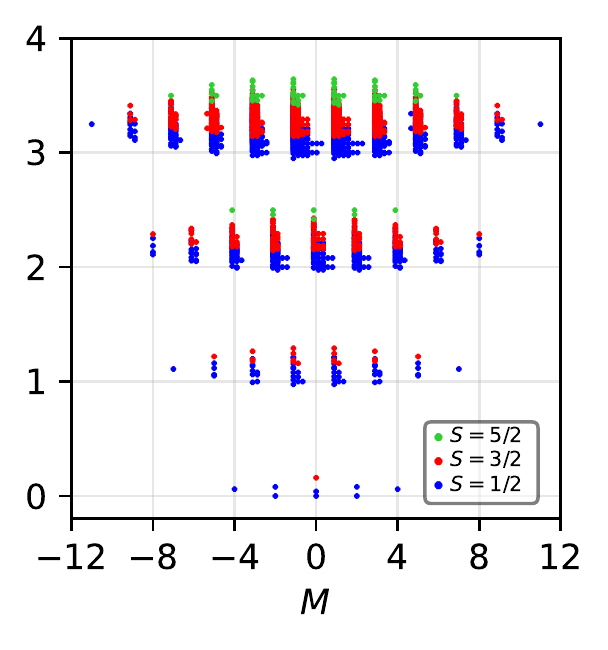}} } \\[-2ex]
\subfigure[Internal breathing modes $(R^\dagger)^a$]
	{\raisebox{-0.2cm}{\includegraphics[scale=0.8]{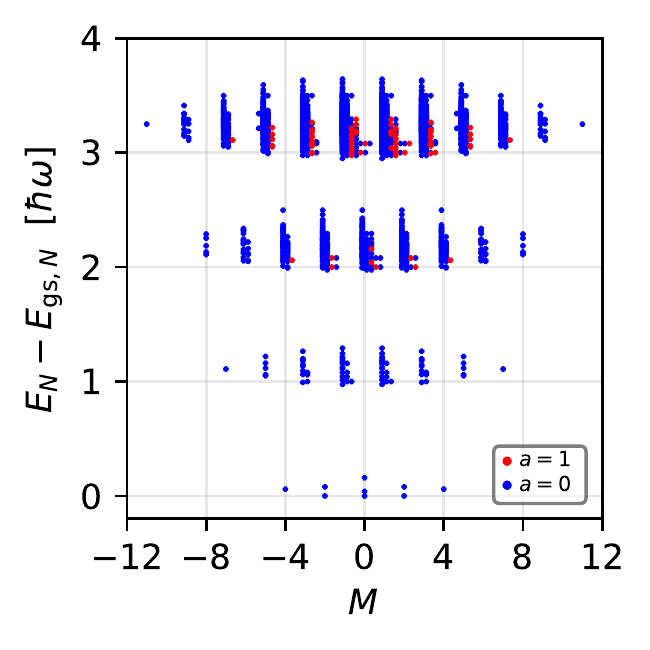}} }
\subfigure[Center-of-mass excitations $(Q_+^\dagger)^b$]
	{\raisebox{-0.2cm}{\includegraphics[scale=0.8]{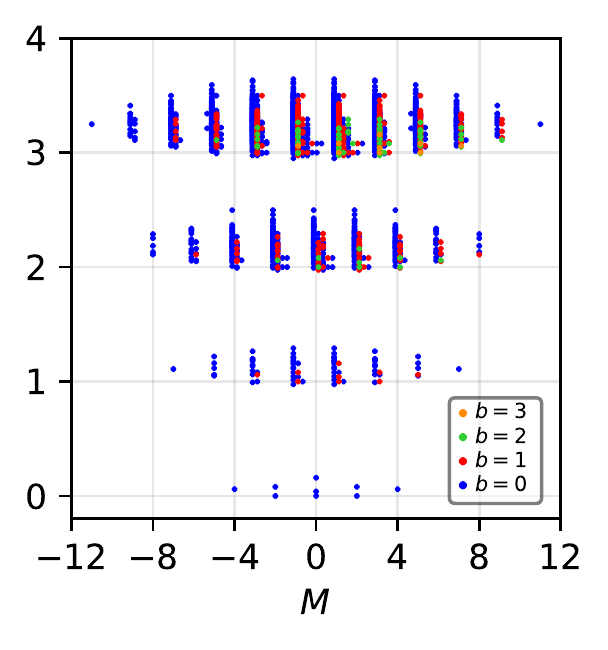}} }
\subfigure[Center-of-mass excitations $(Q_-^\dagger)^c$]
	{\raisebox{-0.2cm}{\includegraphics[scale=0.8]{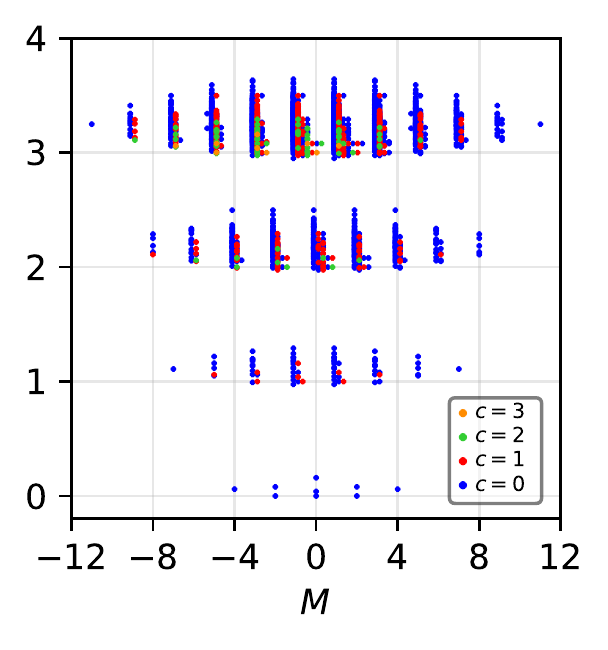}} }
	\caption{
Excitations energies for $N=9$ particles in a harmonic trap ordered by angular momentum for an attractive interaction \mbox{$g=-1$}. Data points are the same for all panels, but the color coding differs: States are grouped according to (a) primary and nonprimary states, (b) Casimir operator $\langle T \rangle$, (c) total spin quantum number $\langle S^2 \rangle = s (s+1)$, (d) Number of internal breathing mode excitations $a$, and (e) and (f) number of center-of-mass excitations $b$ and $c$.
}
\label{fig:S4}
\end{figure*}
%++++++++++++++++++++++++++++++++++++++++

%++++++++++++++++++++++++++++++++++++++++
\begin{figure*}
\subfigure[primary/nonprimary states]
	{\raisebox{-0.2cm}{\includegraphics[scale=0.8]{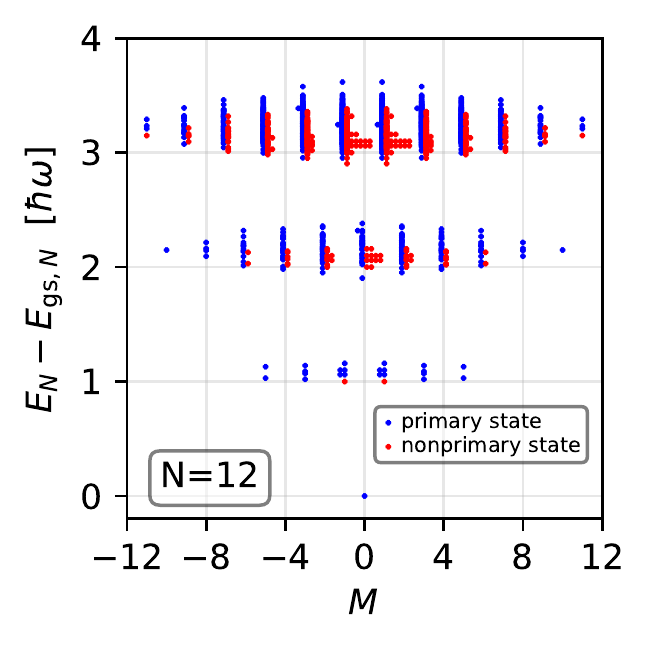}} } 
\subfigure[Casimir $\langle T \rangle$]
	{\raisebox{-0.2cm}{\includegraphics[scale=0.8]{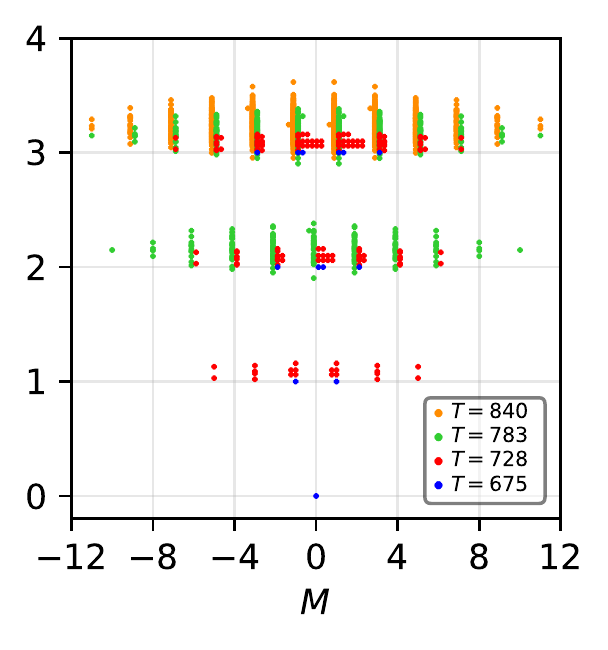}} }
\subfigure[total spin $\langle S^2 \rangle = s (s+1)$]
	{\raisebox{-0.2cm}{\includegraphics[scale=0.8]{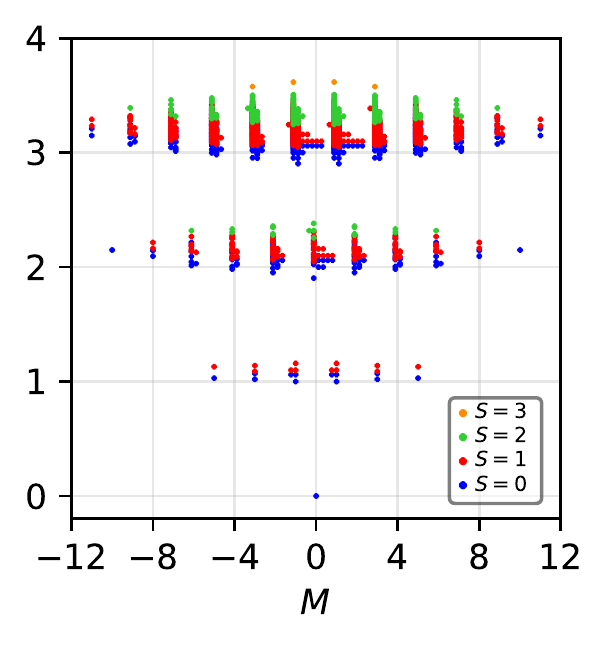}} } \\[-2ex]
\subfigure[Internal breathing modes $(R^\dagger)^a$]
	{\raisebox{-0.2cm}{\includegraphics[scale=0.8]{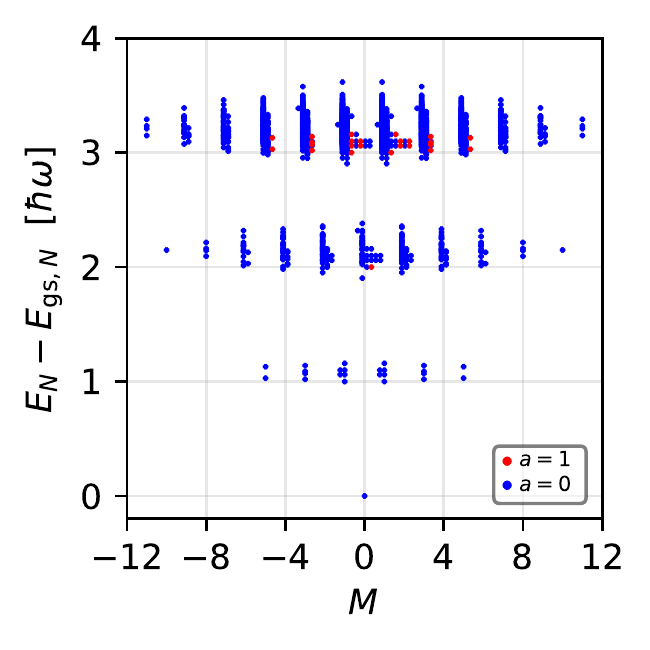}} }
\subfigure[Center-of-mass excitations $(Q_+^\dagger)^b$]
	{\raisebox{-0.2cm}{\includegraphics[scale=0.8]{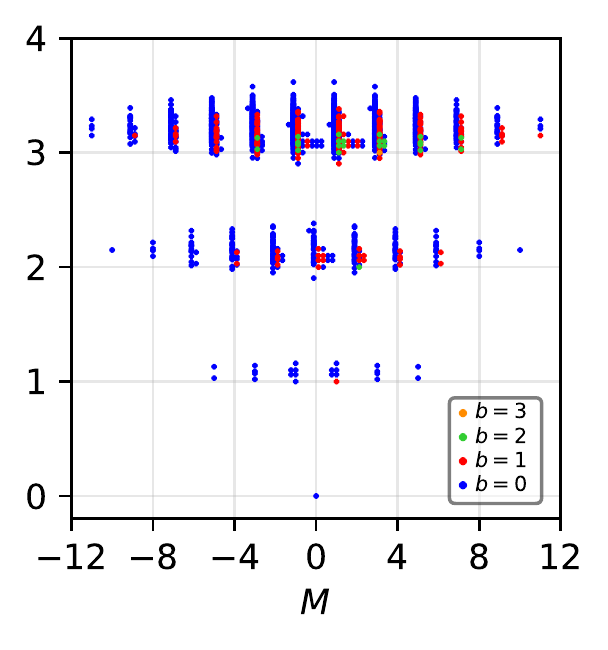}} }
\subfigure[Center-of-mass excitations $(Q_-^\dagger)^c$]
	{\raisebox{-0.2cm}{\includegraphics[scale=0.8]{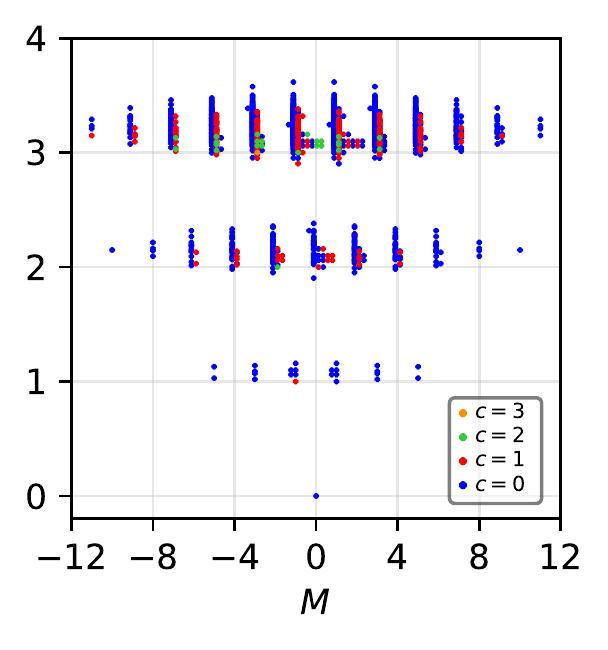}} }
	\caption{
Excitations energies for $N=12$ particles in a harmonic trap ordered by angular momentum for an attractive interaction \mbox{$g=-1$}. Data points are the same for all panels, but the color coding differs: States are grouped according to (a) primary and nonprimary states, (b) Casimir operator $\langle T \rangle$, (c) total spin quantum number $s$, (d) Number of internal breathing mode excitations $a$, and (e) and (f) number of center-of-mass excitations $b$ and $c$.
}
\label{fig:S5}
\end{figure*}
%++++++++++++++++++++++++++++++++++++++++

\clearpage

\begin{table}[h]
\caption{\label{tab:ex} Primary states for \mbox{$N=2$} up to the fourth excitation level. For $|M|\neq0$, there are two primary states with opposite angular momentum $\pm M$ and equal eigenenergy $E_2^{(1)}$ ($\hbar \omega=1$), spin $S$, and Casimir $T$. There is one primary state in the nondegenerate ground state with two new primary states at every excited level.}
\begin{ruledtabular}
\begin{tabular}{lllll  llllll  llllll  llllll}
$lvl$ & $-E_{2}^{(1)}$ & S & $|M|$ &T & \quad\quad\quad\quad\quad & 
$lvl$ & $-E_{2}^{(1)}$ & S & $|M|$ & T & \quad\quad\quad\quad\quad & 
$lvl$ & $-E_{2}^{(1)}$ & S & $|M|$ &T & \quad\quad\quad\quad\quad & 
$lvl$ & $-E_{2}^{(1)}$ & S & $|M|$ &T  \\ 
         0 & -0.15915 & 0 & 0 & -1 &&
          2 & 0 & 0 & 2 & 3 &&
           3 & 0 & 1 & 3 & 8 && 
           4 & 0 & 0 & 4 & 15 \\ 
         1 & 0 & 1 & 1 & 0 
\end{tabular}
\end{ruledtabular}
\end{table}

\begin{table}[h]
\caption{\label{tab:ex} Primary states for \mbox{$N=6$} up to the second excitation level. For $|M|\neq0$, there are two primary states with opposite angular momentum $\pm M$ and equal eigenenergy $E_6^{(1)}$ ($\hbar \omega=1$), spin $S$, and Casimir $T$. There is one primary state in the nondegenerate ground state, $10$ new primary states at the first excited level, 40 at the second excited level, 106 at the third excited level, and 265 at the fourth excited level.}
\begin{ruledtabular}
\begin{tabular}{llllll | llllll | llllll | lllll}
$lvl$ & $-E_{6}^{(1)}/g$  & S \quad\quad& $|M|$ &T & \qquad & 
$lvl$ & $-E_{6}^{(1)}/g$  & S \quad\quad & $|M|$ & T & \qquad & 
$lvl$ & $-E_{6}^{(1)}/g$  & S \quad\quad & $|M|$ & T & \qquad & 
$lvl$ & $-E_{6}^{(1)}/g$  & S \quad\quad & $|M|$ &T  \\ 
        0 & -0.79577 & 0 & 0 & 63 && 2 & -0.74814 & 0 & 4 & 99 && 2 & -0.65651 & 0 & 0 & 99 && 2 & -0.54500 & 1 & 0 & 99 \\
        1 & -0.77588 & 0 & 3 & 80 && 2 & -0.74097 & 0 & 0 & 99 && 2 & -0.64808 & 1 & 2 & 99 && 2 & -0.53715 & 1 & 4 & 99 \\
        1 & -0.73609 & 0 & 1 & 80 && 2 & -0.73189 & 0 & 2 & 99 && 2 & -0.61673 & 1 & 4 & 99 && 2 & -0.53065 & 0 & 2 & 99 \\
        1 & -0.69630 & 1 & 1 & 80 && 2 & -0.70406 & 1 & 2 & 99 && 2 & -0.61262 & 0 & 2 & 99 && 2 & -0.52740 & 1 & 2 & 99 \\
        1 & -0.65651 & 1 & 3 & 80 && 2 & -0.69630 & 1 & 0 & 99 && 2 & -0.60468 & 0 & 4 & 99 && 2 & -0.48995 & 1 & 0 & 99 \\
        1 & -0.63662 & 1 & 1 & 80 && 2 & -0.69299 & 0 & 2 & 99 && 2 & -0.60424 & 1 & 0 & 99 && 2 & -0.43768 & 2 & 2 & 99 \\ 
        2 & -0.87285 & 0 & 0 & 99 && 2 & -0.68846 & 1 & 0 & 99 && 2 & -0.59446 & 0 & 0 & 99 && 2 & -0.39789 & 2 & 0 & 99 \\
        2 & -0.83378 & 0 & 2 & 99 && 2 & -0.65651 & 0 & 6 & 99 && 2 & -0.54757 & 1 & 2 & 99 &&  &  &  &  &\\ 
\end{tabular}
\end{ruledtabular}
\end{table}

\begin{table}[h]
\caption{\label{tab:ex} Primary states for \mbox{$N=9$} up to the second excitation level. For $|M|\neq0$, there are two primary states with opposite angular momentum $\pm M$ and equal eigenenergy $E_9^{(1)}$ ($\hbar \omega=1$), spin $S$, and Casimir $T$. There are 9 primary state in the degenerate ground state with 54 new primary states at the first excited level, 233 at the second excited level, and 768 at the third excited level.}
\begin{ruledtabular}
\begin{tabular}{llllll | llllll | llllll | lllll}
$lvl$ & $-E_{9}^{(1)}/g$  & S \quad\quad& $|M|$ &T & \qquad & 
$lvl$ & $-E_{9}^{(1)}/g$  & S \quad\quad & $|M|$ & T & \qquad & 
$lvl$ & $-E_{9}^{(1)}/g$  & S \quad\quad & $|M|$ & T & \qquad & 
$lvl$ & $-E_{9}^{(1)}/g$  & S \quad\quad & $|M|$ &T  \\ 
        0 & -1.43239 & 1/2 & $$2 & 288 &  & 2 & -1.37291 & 1/2 & $$2 & 360 &  & 2 & -1.25280 & 3/2 & 0 & 360 &  & 2 & -1.15288 & 1/2 & $$2 & 360 \\  
        0 & -1.43239 & 1/2 & 0 & 288 &  & 2 & -1.37219 & 1/2 & $$4 & 360 &  & 2 & -1.24991 & 1/2 & $$2 & 360 &  & 2 & -1.14880 & 1/2 & 0 & 360 \\  
        0 & -1.39261 & 1/2 & 0 & 288 &  & 2 & -1.36304 & 1/2 & 0 & 360 &  & 2 & -1.24799 & 1/2 & 0 & 360 &  & 2 & -1.14863 & 1/2 & 0 & 360 \\  
        0 & -1.37271 & 1/2 & $$4 & 288 &  & 2 & -1.35840 & 1/2 & $$4 & 360 &  & 2 & -1.24736 & 3/2 & $$2 & 360 &  & 2 & -1.14575 & 1/2 & 0 & 360 \\  
        0 & -1.35282 & 1/2 & $$2 & 288 &  & 2 & -1.35131 & 1/2 & 0 & 360 &  & 2 & -1.24637 & 1/2 & $$8 & 360 &  & 2 & -1.14393 & 3/2 & $$8 & 360 \\  
        0 & -1.27324 & 3/2 & 0 & 288 &  & 2 & -1.35014 & 1/2 & $$2 & 360 &  & 2 & -1.24200 & 3/2 & $$4 & 360 &  & 2 & -1.14309 & 1/2 & $$2 & 360 \\  
        1 & -1.45704 & 1/2 & $$1 & 323 &  & 2 & -1.34751 & 1/2 & $$6 & 360 &  & 2 & -1.23990 & 3/2 & $$2 & 360 &  & 2 & -1.13917 & 3/2 & $$6 & 360 \\  
        1 & -1.43962 & 1/2 & $$3 & 323 &  & 2 & -1.34201 & 1/2 & 0 & 360 &  & 2 & -1.23843 & 1/2 & $$2 & 360 &  & 2 & -1.13583 & 3/2 & $$4 & 360 \\  
        1 & -1.41815 & 1/2 & $$1 & 323 &  & 2 & -1.33916 & 1/2 & $$4 & 360 &  & 2 & -1.23743 & 3/2 & 0 & 360 &  & 2 & -1.13424 & 3/2 & $$2 & 360 \\  
        1 & -1.38681 & 1/2 & $$1 & 323 &  & 2 & -1.33898 & 1/2 & 0 & 360 &  & 2 & -1.22884 & 1/2 & $$6 & 360 &  & 2 & -1.13137 & 3/2 & 0 & 360 \\  
        1 & -1.38187 & 1/2 & $$5 & 323 &  & 2 & -1.33846 & 1/2 & $$2 & 360 &  & 2 & -1.22816 & 1/2 & $$4 & 360 &  & 2 & -1.12651 & 1/2 & $$2 & 360 \\  
        1 & -1.37086 & 1/2 & $$3 & 323 &  & 2 & -1.33712 & 1/2 & $$2 & 360 &  & 2 & -1.22779 & 3/2 & $$6 & 360 &  & 2 & -1.12518 & 3/2 & $$2 & 360 \\  
        1 & -1.35282 & 1/2 & $$1 & 323 &  & 2 & -1.33054 & 1/2 & 0 & 360 &  & 2 & -1.22753 & 1/2 & $$2 & 360 &  & 2 & -1.12156 & 3/2 & 0 & 360 \\  
        1 & -1.33854 & 1/2 & $$3 & 323 &  & 2 & -1.32634 & 1/2 & $$2 & 360 &  & 2 & -1.22509 & 1/2 & 0 & 360 &  & 2 & -1.12017 & 1/2 & $$4 & 360 \\  
        1 & -1.32298 & 1/2 & $$7 & 323 &  & 2 & -1.32261 & 1/2 & $$4 & 360 &  & 2 & -1.21989 & 1/2 & $$4 & 360 &  & 2 & -1.11837 & 3/2 & $$4 & 360 \\  
        1 & -1.31752 & 1/2 & $$1 & 323 &  & 2 & -1.32159 & 1/2 & 0 & 360 &  & 2 & -1.21933 & 1/2 & 0 & 360 &  & 2 & -1.11444 & 1/2 & $$6 & 360 \\  
        1 & -1.31512 & 1/2 & $$5 & 323 &  & 2 & -1.32087 & 1/2 & $$2 & 360 &  & 2 & -1.21832 & 3/2 & $$4 & 360 &  & 2 & -1.11299 & 3/2 & $$2 & 360 \\  
        1 & -1.31102 & 1/2 & $$1 & 323 &  & 2 & -1.31279 & 1/2 & 0 & 360 &  & 2 & -1.21617 & 3/2 & $$2 & 360 &  & 2 & -1.10824 & 3/2 & $$6 & 360 \\  
        1 & -1.29761 & 1/2 & $$3 & 323 &  & 2 & -1.31102 & 1/2 & $$6 & 360 &  & 2 & -1.21464 & 1/2 & $$2 & 360 &  & 2 & -1.10689 & 1/2 & 0 & 360 \\  
        1 & -1.28739 & 1/2 & $$3 & 323 &  & 2 & -1.31090 & 1/2 & 0 & 360 &  & 2 & -1.21200 & 3/2 & 0 & 360 &  & 2 & -1.10484 & 3/2 & $$2 & 360 \\  
        1 & -1.27843 & 1/2 & $$1 & 323 &  & 2 & -1.30806 & 1/2 & $$2 & 360 &  & 2 & -1.20865 & 1/2 & $$6 & 360 &  & 2 & -1.09934 & 3/2 & $$4 & 360 \\  
        1 & -1.27194 & 1/2 & $$5 & 323 &  & 2 & -1.30302 & 1/2 & $$6 & 360 &  & 2 & -1.20386 & 3/2 & $$2 & 360 &  & 2 & -1.09558 & 3/2 & $$6 & 360 \\  
        1 & -1.25937 & 3/2 & $$1 & 323 &  & 2 & -1.30293 & 1/2 & $$4 & 360 &  & 2 & -1.20223 & 3/2 & $$4 & 360 &  & 2 & -1.09449 & 1/2 & $$4 & 360 \\  
        1 & -1.25335 & 3/2 & $$3 & 323 &  & 2 & -1.30110 & 1/2 & $$8 & 360 &  & 2 & -1.20211 & 1/2 & $$4 & 360 &  & 2 & -1.08835 & 3/2 & 0 & 360 \\  
        1 & -1.24630 & 3/2 & $$1 & 323 &  & 2 & -1.29897 & 1/2 & $$2 & 360 &  & 2 & -1.20091 & 1/2 & 0 & 360 &  & 2 & -1.08585 & 3/2 & $$4 & 360 \\  
        1 & -1.23947 & 3/2 & $$3 & 323 &  & 2 & -1.29284 & 1/2 & $$4 & 360 &  & 2 & -1.19952 & 1/2 & 0 & 360 &  & 2 & -1.08356 & 1/2 & $$2 & 360 \\  
        1 & -1.23369 & 1/2 & $$3 & 323 &  & 2 & -1.29166 & 1/2 & 0 & 360 &  & 2 & -1.19769 & 3/2 & 0 & 360 &  & 2 & -1.08143 & 1/2 & 0 & 360 \\  
        1 & -1.22474 & 1/2 & $$1 & 323 &  & 2 & -1.28680 & 3/2 & $$2 & 360 &  & 2 & -1.19609 & 1/2 & $$6 & 360 &  & 2 & -1.07947 & 3/2 & 0 & 360 \\  
        1 & -1.21994 & 1/2 & $$1 & 323 &  & 2 & -1.28562 & 1/2 & $$2 & 360 &  & 2 & -1.19550 & 3/2 & $$2 & 360 &  & 2 & -1.07350 & 3/2 & $$2 & 360 \\  
        1 & -1.21356 & 3/2 & $$5 & 323 &  & 2 & -1.27847 & 1/2 & $$6 & 360 &  & 2 & -1.18969 & 1/2 & $$2 & 360 &  & 2 & -1.06951 & 3/2 & 0 & 360 \\  
        1 & -1.18764 & 3/2 & $$1 & 323 &  & 2 & -1.27820 & 3/2 & 0 & 360 &  & 2 & -1.18864 & 3/2 & $$6 & 360 &  & 2 & -1.06441 & 3/2 & $$4 & 360 \\  
        1 & -1.16774 & 3/2 & $$3 & 323 &  & 2 & -1.27622 & 1/2 & $$2 & 360 &  & 2 & -1.18664 & 1/2 & 0 & 360 &  & 2 & -1.05649 & 1/2 & 0 & 360 \\  
        1 & -1.14103 & 3/2 & $$1 & 323 &  & 2 & -1.27266 & 3/2 & 0 & 360 &  & 2 & -1.18610 & 1/2 & 0 & 360 &  & 2 & -1.05201 & 3/2 & 0 & 360 \\  
        2 & -1.43950 & 1/2 & 0 & 360 &  & 2 & -1.27232 & 1/2 & $$4 & 360 &  & 2 & -1.18381 & 3/2 & $$2 & 360 &  & 2 & -1.04780 & 3/2 & $$2 & 360 \\  
        2 & -1.43947 & 1/2 & $$2 & 360 &  & 2 & -1.27044 & 1/2 & 0 & 360 &  & 2 & -1.18024 & 1/2 & $$8 & 360 &  & 2 & -1.03136 & 3/2 & 0 & 360 \\  
        2 & -1.43440 & 1/2 & 0 & 360 &  & 2 & -1.26987 & 3/2 & $$2 & 360 &  & 2 & -1.17752 & 3/2 & $$2 & 360 &  & 2 & -1.02862 & 3/2 & $$2 & 360 \\  
        2 & -1.42601 & 1/2 & $$2 & 360 &  & 2 & -1.26970 & 1/2 & 0 & 360 &  & 2 & -1.17749 & 3/2 & $$4 & 360 &  & 2 & -1.01762 & 3/2 & $$2 & 360 \\  
        2 & -1.42340 & 1/2 & $$4 & 360 &  & 2 & -1.26920 & 1/2 & $$2 & 360 &  & 2 & -1.17617 & 1/2 & $$4 & 360 &  & 2 & -1.01461 & 5/2 & 0 & 360 \\  
        2 & -1.41362 & 1/2 & $$2 & 360 &  & 2 & -1.26388 & 1/2 & 0 & 360 &  & 2 & -1.17570 & 3/2 & 0 & 360 &  & 2 & -1.00593 & 3/2 & 0 & 360 \\  
        2 & -1.40214 & 1/2 & 0 & 360 &  & 2 & -1.26260 & 1/2 & $$4 & 360 &  & 2 & -1.17483 & 1/2 & $$2 & 360 &  & 2 & -0.97482 & 5/2 & $$2 & 360 \\  
        2 & -1.38782 & 1/2 & $$4 & 360 &  & 2 & -1.25991 & 1/2 & 0 & 360 &  & 2 & -1.17296 & 3/2 & 0 & 360 &  & 2 & -0.93504 & 5/2 & 0 & 360 \\  
        2 & -1.38769 & 1/2 & 0 & 360 &  & 2 & -1.25675 & 3/2 & 0 & 360 &  & 2 & -1.17240 & 1/2 & $$2 & 360 &  & 2 & -0.93504 & 5/2 & $$2 & 360 \\  
        2 & -1.38269 & 1/2 & $$2 & 360 &  & 2 & -1.25655 & 1/2 & $$4 & 360 &  & 2 & -1.15718 & 1/2 & $$4 & 360 &  & 2 & -0.93504 & 5/2 & $$4 & 360 \\  
        2 & -1.37661 & 1/2 & $$6 & 360 &  & 2 & -1.25459 & 3/2 & $$4 & 360 &  &  &  &  &  &  &  &  &  &  &  &  \\  
\end{tabular}
\end{ruledtabular}
\end{table}

\begin{table}[H]
\caption{\label{tab:ex} Primary states for \mbox{$N=12$} up to the second excitation level. For $|M|\neq0$, there are two primary states with opposite angular momentum $\pm M$ and equal eigenenergy $E_{12}^{(1)}$ ($\hbar \omega=1$), spin $S$, and Casimir $T$. There is one primary state in the nondegenerate ground state with 22 new primary states at the first excited level, 178 at the second excited level, and 798 at the third excited level.}
\begin{ruledtabular}
\begin{tabular}{llllll | llllll | llllll | lllll}
$lvl$ & $-E_{12}^{(1)}/g$  & S \quad\quad& $|M|$ &T & \qquad & 
$lvl$ & $-E_{12}^{(1)}/g$  & S \quad\quad & $|M|$ & T & \qquad & 
$lvl$ & $-E_{12}^{(1)}/g$  & S \quad\quad & $|M|$ & T & \qquad & 
$lvl$ & $-E_{12}^{(1)}/g$  & S \quad\quad & $|M|$ &T  \\ 
        0 & -2.22817 & 0 & 0 & 675 && 2 & -2.15872 & 1 & 2 & 783 && 2 & -2.07225 & 1 & 6 & 783 && 2 & -2.00765 & 0 & 2 & 783 \\ 
        1 & -2.20827 & 0 & 3 & 728 && 2 & -2.15538 & 1 & 4 & 783 && 2 & -2.07134 & 1 & 0 & 783 && 2 & -1.99990 & 1 & 0 & 783 \\ 
        1 & -2.19833 & 0 & 5 & 728 && 2 & -2.14907 & 1 & 0 & 783 && 2 & -2.07072 & 0 & 2 & 783 && 2 & -1.99607 & 1 & 2 & 783 \\ 
        1 & -2.16849 & 0 & 1 & 728 && 2 & -2.14399 & 0 & 0 & 783 && 2 & -2.07033 & 0 & 0 & 783 && 2 & -1.99492 & 0 & 0 & 783 \\ 
        1 & -2.16849 & 0 & 1 & 728 && 2 & -2.14222 & 0 & 4 & 783 && 2 & -2.06549 & 1 & 8 & 783 && 2 & -1.99471 & 0 & 0 & 783 \\ 
        1 & -2.15854 & 0 & 3 & 728 && 2 & -2.13840 & 0 & 2 & 783 && 2 & -2.06456 & 1 & 2 & 783 && 2 & -1.99308 & 1 & 2 & 783 \\ 
        1 & -2.13864 & 1 & 3 & 728 && 2 & -2.13360 & 0 & 6 & 783 && 2 & -2.06311 & 0 & 0 & 783 && 2 & -1.98009 & 1 & 4 & 783 \\
        1 & -2.12870 & 1 & 1 & 728 && 2 & -2.13317 & 0 & 8 & 783 && 2 & -2.06051 & 1 & 4 & 783 && 2 & -1.97684 & 0 & 4 & 783 \\
        1 & -2.12870 & 1 & 1 & 728 && 2 & -2.13292 & 0 & 2 & 783 && 2 & -2.05672 & 1 & 2 & 783 && 2 & -1.97277 & 2 & 0 & 783 \\
        1 & -2.09886 & 1 & 5 & 728 && 2 & -2.12990 & 1 & 4 & 783 && 2 & -2.05632 & 0 & 2 & 783 && 2 & -1.97126 & 1 & 2 & 783 \\ 
        1 & -2.08891 & 1 & 3 & 728 && 2 & -2.12649 & 0 & 0 & 783 && 2 & -2.04718 & 1 & 2 & 783 && 2 & -1.96926 & 1 & 0 & 783 \\ 
        1 & -2.06901 & 1 & 1 & 728 && 2 & -2.12243 & 1 & 0 & 783 && 2 & -2.04511 & 0 & 6 & 783 && 2 & -1.96298 & 1 & 4 & 783 \\ 
        2 & -2.32493 & 0 & 0 & 783 && 2 & -2.12180 & 1 & 2 & 783 && 2 & -2.04189 & 1 & 6 & 783 && 2 & -1.96184 & 1 & 2 & 783 \\
        2 & -2.27684 & 0 & 2 & 783 && 2 & -2.11758 & 0 & 2 & 783 && 2 & -2.04005 & 1 & 0 & 783 && 2 & -1.96104 & 1 & 6 & 783 \\
        2 & -2.24575 & 0 & 4 & 783 && 2 & -2.11633 & 1 & 0 & 783 && 2 & -2.03895 & 1 & 0 & 783 && 2 & -1.95701 & 1 & 0 & 783 \\ 
        2 & -2.23689 & 0 & 2 & 783 && 2 & -2.11329 & 0 & 0 & 783 && 2 & -2.03801 & 1 & 4 & 783 && 2 & -1.95346 & 1 & 2 & 783 \\ 
        2 & -2.22320 & 0 & 4 & 783 && 2 & -2.11273 & 1 & 2 & 783 && 2 & -2.03221 & 0 & 4 & 783 && 2 & -1.95317 & 1 & 4 & 783 \\ 
        2 & -2.21493 & 0 & 6 & 783 && 2 & -2.10358 & 0 & 4 & 783 && 2 & -2.03128 & 1 & 6 & 783 && 2 & -1.94965 & 2 & 2 & 783 \\
        2 & -2.20556 & 0 & 0 & 783 && 2 & -2.09352 & 1 & 2 & 783 && 2 & -2.03049 & 0 & 2 & 783 && 2 & -1.94719 & 0 & 2 & 783 \\ 
        2 & -2.19335 & 0 & 2 & 783 && 2 & -2.09190 & 1 & 6 & 783 && 2 & -2.02912 & 1 & 4 & 783 && 2 & -1.93799 & 2 & 2 & 783 \\ 
        2 & -2.19210 & 0 & 0 & 783 && 2 & -2.09125 & 0 & 4 & 783 && 2 & -2.02852 & 1 & 0 & 783 && 2 & -1.92393 & 2 & 4 & 783 \\ 
        2 & -2.18431 & 0 & 0 & 783 && 2 & -2.09014 & 0 & 6 & 783 && 2 & -2.02720 & 0 & 0 & 783 && 2 & -1.92265 & 1 & 0 & 783 \\
        2 & -2.18325 & 0 & 6 & 783 && 2 & -2.08609 & 0 & 2 & 783 && 2 & -2.02019 & 1 & 0 & 783 && 2 & -1.90986 & 2 & 6 & 783 \\ 
        2 & -2.18239 & 0 & 0 & 783 && 2 & -2.08482 & 1 & 4 & 783 && 2 & -2.01765 & 1 & 0 & 783 && 2 & -1.90986 & 2 & 0 & 783 \\
        2 & -2.18070 & 0 & 2 & 783 && 2 & -2.08444 & 0 & 8 & 783 && 2 & -2.01744 & 1 & 4 & 783 && 2 & -1.90986 & 2 & 0 & 783 \\
        2 & -2.17993 & 1 & 0 & 783 && 2 & -2.07896 & 0 & 10 & 783 && 2 & -2.01320 & 1 & 2 & 783 && 2 & -1.89579 & 2 & 4 & 783 \\ 
        2 & -2.16995 & 0 & 2 & 783 && 2 & -2.07746 & 1 & 0 & 783 && 2 & -2.01314 & 0 & 6 & 783 && 2 & -1.88172 & 2 & 2 & 783 \\
        2 & -2.16472 & 1 & 0 & 783 && 2 & -2.07590 & 0 & 4 & 783 && 2 & -2.01285 & 1 & 8 & 783 && 2 & -1.87007 & 2 & 2 & 783 \\
        2 & -2.16446 & 1 & 2 & 783 && 2 & -2.07546 & 1 & 2 & 783 && 2 & -2.00933 & 1 & 0 & 783 && 2 & -1.84695 & 2 & 0 & 783 \\ 
        2 & -2.16288 & 0 & 4 & 783 && 2 & -2.07420 & 1 & 4 & 783 &&  &  &  &  &  &&  &  &  &  &  
\end{tabular}
\end{ruledtabular}
\end{table}

\section{Scaling of primary states}

\begin{figure*}[h]
\subfigure[]
	{\raisebox{-0.2cm}{\includegraphics[scale=1]{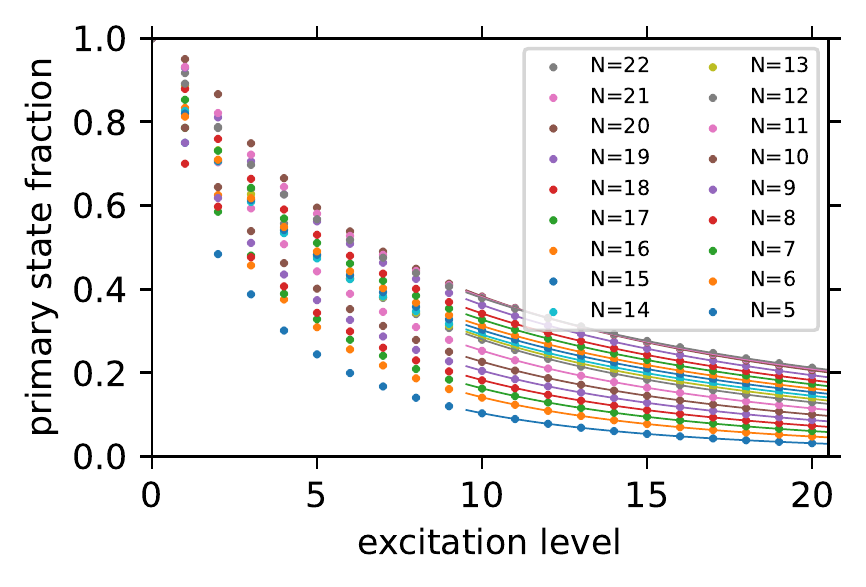}}\qquad\quad } \\[4ex]
\subfigure[] {
	\raisebox{-0.2cm}{\includegraphics[scale=0.8]{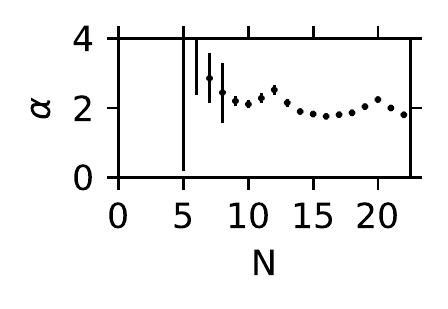}}
	\raisebox{-0.2cm}{\includegraphics[scale=0.8]{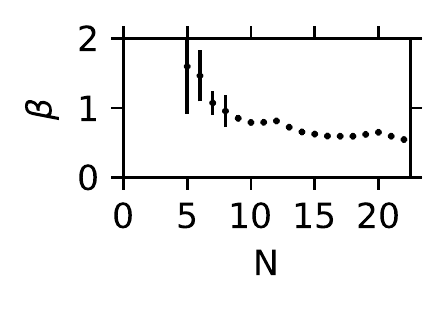}} 
	\raisebox{-0.2cm}{\includegraphics[scale=0.8]{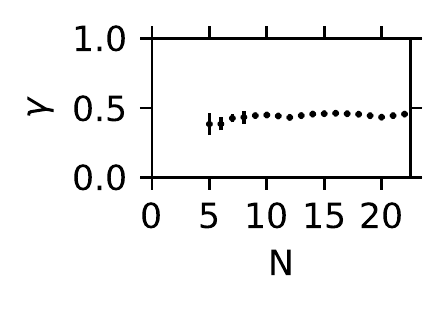}} \quad \quad}
	\caption{
(a) Fraction of primary states for different particle numbers $N=5-22$ as a function of excitation level. Even at large excitation levels, there is a sizable fraction of primary states. (b) Fit parameters for an exponential fit of the form $\alpha \exp[- \beta l^\gamma]$ to the primary state fraction as a function of excitation level.
}
\label{fig:S6}
\end{figure*}

Figure~3 of the main text as well as Figs.~\ref{fig:S2}-\ref{fig:S5} show a trend where the fraction of primary states within an excitation level increases with both excitation level and with particle number. To quantify this, we show in in Fig.~\ref{fig:S6}(a) the ratio of primary states to the dimension of the degenerate level Hilbert space up to an excitation level $20$ for different particle numbers $N=5-22$. The number of primary states per level can be determined recursively from the Hilbert space dimension by noting that primary states at one level generate $\lfloor \frac{l^2}{4} \rfloor$ derived nonprimary states at $l$ levels higher (see Fig.~2 and the discussion in the main text). As is apparent from the figure, the fraction of primary states decreases with the excitation level, albeit quite slowly, and increases with particle number for a given level. \\

For large excitation levels, we can perform an exponential fit to the data of the form $\alpha \exp[- \beta l^\gamma]$, where $l$ is the excitation level and $\alpha, \beta, \gamma$ are fit parameters that depend on the particle number $N$. Figure~\ref{fig:S6}(b) shows results for these parameters for a fit that includes the largest 10 excitation levels as a function of particle number, which are indicated by the continuous lines in Fig.~\ref{fig:S6}(a). The fit is reliable for larger particle numbers $N\geq 10$ with a weak dependence on $N$. In addition, the fit parameters show a weak shell effect with kinks at the magic number configurations $N=12$ and $N=20$.

%++++++++++++++++++++++++++++++++++++++++
%++++++++++++++++++++++++++++++++++++++++

\newpage